\documentclass[14pt,a4paper]{article}
\usepackage[utf8x]{inputenc}
\usepackage{jheppub}
\usepackage{rotating}
\usepackage{bm}        
\usepackage{amssymb}   
\usepackage{dsfont}
\usepackage{graphicx}
\usepackage{skak}
\usepackage{cleveref}
\graphicspath{ {./images/} }
\usepackage{tikz}
\usepackage{tikz-feynman}
\usetikzlibrary{arrows.meta}
\renewcommand{\[}{\begin{equation}\begin{aligned}}
\renewcommand{\]}{\end{aligned}\end{equation}}
\def\be{\begin{equation}}
\def\ee{\end{equation}}
\def\beal{\begin{equation}\begin{aligned}}

\let\emptyset\varnothing
\def\eeal{\end{aligned}\end{equation}}
\def\nn{\nonumber}
\newcommand{\waveshape}{\alpha_\eta(k)}

\def\braket#1{\langle #1 \rangle}

\def\lexp{\biggl\langle\!\!\!\biggl\langle}
\def\rexp{\biggr\rangle\!\!\!\biggr\rangle}
\newcommand{\CCut}{\operatorname{CCut}}
\newcommand{\ph}{\hat p}   
\newcommand{\KMOCav}[1]{\lexp #1 \rexp}

\renewcommand{\vec}[1]{\ensuremath{\boldsymbol{#1}}}

\renewcommand{\Re}{\operatorname{Re}}

\newcommand{\dd}{\mathrm{d}}

 \newcommand{\sa}{\mathsf a}
  \newcommand{\sS}{\mathsf S}
\newcommand{\hdelta}{\hat{\delta}}
\newcommand{\hd}{\hat{\mathrm{d}}}

\title{A Spin on Radiation Reaction with Cuts}

\author[1,2]{Asaad Elkhidir,}
\emailAdd{A.E.H.Elkhidir@sms.ed.ac.uk}
\author[3,4]{Matteo Sergola,}
\emailAdd{msergola@physics.ucla.edu}
\author[5]{and Pablo Vives Matasan}
\emailAdd{p.vivesmatasan@qmul.ac.uk}

\affiliation[1]{
Higgs Centre for Theoretical Physics, School of Physics and Astronomy,\\
The University of Edinburgh, Edinburgh EH9 3JZ, Scotland, UK}
\affiliation[2]{
Institut des Hautes Etudes Scientifiques, 91440 Bures-sur-Yvette, France}
\affiliation[3]{
Institut de Physique Théorique, CEA, CNRS, \\Université Paris-Saclay, F–91191 Gif-sur-Yvette cedex, France}
\affiliation[4]{Mani L. Bhaumik Institute for Theoretical Physics, \\
 University of California at Los Angeles, Los Angeles, CA 90095, USA}
\affiliation[5]{Centre for Theoretical Physics and Astronomy, Department of Physics and Astronomy, \\
Queen Mary University of London, Mile End Road, London E1 4NS, UK}

\abstract{
We derive the radiation-reaction equations of motion for a spinning charged body, the $\sqrt{\text{Kerr}}$ particle, through quadratic order in its spin, generalising the Abraham--Lorentz--Dirac equation.  Extending an argument of Coleman, we obtain radiation reaction equations by evaluating the self-field of the spinning body on its own worldline, and removing its divergences with dimensional regularisation. We use the resulting equations of motion to compute the electromagnetic waveform emitted in the scattering of the spinning body off a spectator charge, both at tree level and including the leading radiation-reaction correction, which enters at one loop.
Then, we apply the construction introduced in a previous work by some of the authors -- where radiation reaction is encoded in  unitarity cuts involving  Compton amplitudes -- to spinning particles, and recompute the radiation-reaction contribution to the one-loop waveform. 
We find that the waveform calculated using classical and amplitude methods agree through quadratic order in spin, providing a check on our radiation-reaction equations.  Along the way, we also reproduce known results for linear-in-spin self-force effects and for the spinless self-force in six dimensions.}

\begin{document}

\maketitle

\section{Introduction}

 An accelerated charge radiates, and the radiation carries away energy, momentum and angular momentum. Conservation then requires that the charge feels a recoil from emission. For a structureless point charge this self-force is the Abraham--Lorentz--Dirac (ALD) force \cite{lorentz1892theorie,PlanckUeberES,abraham,Dirac:1938nz}.

For a spinning body, the self-field also couples to its magnetic dipole and higher multipoles, modifying the velocity, momentum, and spin equations of motion  (EOMs) that govern the dynamics of the projectile. The purpose of this work is to explicitly derive the radiation reaction force for spinning bodies. 

To gain intuition let us go back to classical nonspinning electrodynamics for a moment. It is a well-known fact that the Lorentz force equation, which describes the acceleration of a particle under electromagnetic (EM) forces, can be derived from the \emph{total} energy-momentum conservation of the system \cite{Landau:1975pou,Jackson}:  $\partial_\mu T^{\mu\nu}=0$, where $T^{\mu\nu}$ is the stress-energy tensor. However the Lorentz force alone is not enough to account for a body's recoil in response to emission of energy, which appears starting at next-to-next-to-leading order of classical scattering observables \cite{Kosower:2018adc,Saketh:2021sri, Bern:2021xze}. To include these effects the Lorentz force law \emph{has to} be  supplemented with the ALD force. Without the latter energy-momentum would not be conserved, but then how do we explain that the classical equations can indeed be derived from the conservation law $\partial_\mu T^{\mu\nu}=0$? The answer is simple: if the right hand side of the Lorentz force equation
\[\label{lforce}
\frac{\dd p^\mu}{\dd\tau}=\frac{Q}{m} F^{\mu\nu}(x(\tau))p_\nu,
\]
is written in terms of an external Maxwell's field  $F^{\mu\nu}$ (for instance generated by a second charge) then what one  is omitting from \eqref{lforce} is the (divergent) field generated by the particle itself, without it the equation above is well-behaved but incomplete. Put differently, the energy tensor in $\partial_\mu T^{\mu\nu}=0$ is only conserved as a sum of both particles' contributions. One way to get around this problem is to include an additional force term to the RHS of \eqref{lforce}
\[ \label{alddd}
f_{\text{ALD}}^\mu=\frac{Q^2}{6\pi m}
      \left(\frac{\dd^2 p^\mu}{\dd\tau^2}+\,\frac{p^\mu}{m^2}\frac{\dd p}{\dd\tau}\cdot \frac{\dd p}{\dd\tau}\right),
\]
which is indeed the ALD equation mentioned before.

In this paper, we show how this intuition carries over to spinning bodies by considering the $\sqrt{\rm Kerr}$ theory \cite{Monteiro:2014cda, Arkani-Hamed:2019ymq, Menezes:2022tcs, Scheopner:2023rzp,Kim:2024grz, deFirmian:2026mln}, and  derive a novel generalisation of the ALD equation to spinning bodies up to quadratic order in spin. We will study this in terms of two independent methods, whose results will be shown to be equivalent. First, inspired by an article of Sidney Coleman \cite{RM-2820-PR}, we will begin with the regularisation of spinning classical fields using techniques borrowed from quantum field theory (QFT) such as dimensional regularisation (dimreg in short). This will allow us to determine a novel set of spinning  equations of motion to describe both conservative and dissipative effects. For instance, we will find that the  equation that generalises \eqref{alddd} to the case where the particle has four-velocity $u^\mu(\tau)$ and spin $\sa^\mu(\tau)$ is
\[\label{mainn}
f^\mu_{\text{ALD},\sa}&= \frac{Q^2}{6\pi}(\ddot u^\mu+\dot{u}^2u^\mu)
+\frac{Q^2}{6\pi}\left(\epsilon^\mu(u,\dot{\sa},\ddot{u})-\epsilon^\mu(\sa,\dot{u},\ddot{u})-2u^\mu\epsilon(\sa,u,\dot{u},\ddot{u})\right)\\& - \frac{Q^2}{120\pi}\Bigl( 5\bigl(11\dot\sa\cdot \dddot u + 16\ddot\sa\cdot\ddot u+10\dddot\sa\cdot\dot u\bigr)\sa^\mu +15\sa\cdot\dddot u\dot\sa^\mu + 20\sa\cdot \ddot u\ddot\sa^\mu
\\&
\qquad\,\,\,\,\,\,\,\,\,\,\,\,\,\,\, +10\sa\cdot\dot u\dddot\sa^\mu - \bigl(10(\sa\cdot\ddot u)^2+\sa^2(49\dot u\cdot \dddot u+33\dot u^2)\bigr)u^\mu - 60\sa\cdot\dddot\sa\dot u^\mu\\
    &\qquad \,\,\,\,\,\,\,\,\,\,\,\,\,\,\,\,\,\,\,\,\,\,\,\,-60\sa\cdot\ddot\sa \ddot u^\mu - 30\sa\cdot\dot\sa \dddot u^\mu+(3\mathsf a^\mu\mathsf a^\nu-\mathsf a^2\eta^{\mu\nu})\,\ddddot u_{\nu}-20 u^\mu \,\mathsf a\cdot\ddddot{\mathsf a}\Bigr)+\mathcal{O}(\sa^3), 
\]
where $\epsilon_\mu(a,b,c)= \epsilon_{\mu\alpha\beta\gamma}a^\alpha b^\beta c^\gamma$ indicates a vector built from the Levi-Civita pseudotensor. We will use these equations to compute the leading-order radiation reaction effects on a spinning observable: the one-loop EM waveform \cite{Elkhidir:2023dco}. 
Then, we will be able to check our results by extending the methods of \cite{Elkhidir:2023dco} with spinning degrees of freedom. In that paper, dissipation\footnote{In this paper we will use the wording ``dissipative" as a synonym of radiation reaction, radiation caused by a particle's own field. ``Conservative" radiation can be similarly defined as radiation generated by an accelerated particle under the influence of an external field. See also \cite{Elkhidir:2023dco}.} is determined via a precise set of unitarity cuts of amplitudes expanded in the classical limit as $\hbar\to 0$. We will show how spin and dissipation can be included straightforwardly and then also compute the EM waveform using on-shell amplitude methods. This framework is deeply rooted in the KMOC framework of \cite{Kosower:2018adc,Maybee:2019jus,Cristofoli:2021vyo}. Both of these methods (classical and amplitude cuts) can in principle be applied to dissipative effects of  generic multipoles as well as  higher orders in spin. However,  for somewhat independent reasons due to spurious poles of classical Compton amplitudes \cite{Arkani-Hamed:2017jhn} (explained in more detail below) we will have to restrict to spin-squared order. This is already enough to both provide new results and non-trivially test our findings. 

Let us finally mention that we are certainly not the first to link spin and dissipation through QFT. In fact, 
the literature contains many calculations of radiation reaction from amplitudes, both at the level of equations of motion and at the level of observables. See \cite{Corben:1941tgm,Ward_1965,ellis1966electromagnetic,Villarroel:1975ij,Teitelboim:1979px} for earlier characterisations of  radiation reaction and spin multipoles, \cite{vanHolten:1990we,Krivitskii:1991,Higuchi:2002qc,Kazinski:2006dh,Kosyakov:2018wek} for more recent derivations using QED Feynman diagrams, \cite{Ilderton:2013tb,Torgrimsson:2021wcj} for advances within strong-field QED and \cite{Jakobsen:2023tvm} for a recent worldline-QFT approach. However,  different authors systematically disregard total-derivative Schott terms \cite{schott1912electromagnetic} that are essential for the calculation of the angular momentum flux as noted in \cite{Bini:2022wrq}, or for the waveform as described in \cite{Elkhidir:2023dco}. Furthermore, dropping total derivatives obviously makes the comparison of different sets of EOMs complicated, if not impossible, see the comments in \cite{Jakobsen:2023tvm}. 
In our work we will go beyond this drawback: we will compare linear-in-spin EOMs to \cite{Jakobsen:2023tvm} but also compute (and check) the on-shell gauge invariant sector of the LO dissipative EM waveform, as well as going to higher spinning multipoles. 

Our paper is divided into sections. After reviewing the main needed tools about radiation reaction and spin in section \ref{general}, we move to the characterisation of spinning radiation reaction in section \ref{sec: crr}. After that, we repeat and validate our results using cuts and scattering amplitudes in section \ref{sec:ccut}. Then we add a small discussion on gravity and conclude in section \ref{sec:concle}. Furthermore, in appendix \ref{appendix subl} we compare to the results of \cite{Jakobsen:2023tvm}.

\subsection*{Miscellaneous  conventions}
We work with  mostly negative signature $(+,-,-,-)$. Factors of $2\pi$ are absorbed with the help of $\hat{\dd}^n p= \dd^n p/(2\pi)^n$, $\hat{\delta}^n(p) =(2\pi)^n \delta ^n (p)$ and we write phase space integrations as 
$$
  \dd\Phi(p_1,\cdots p_n)\equiv \mu^{n(4-D)}\prod_{i=1}^n\hat \dd^D p_i\,\Theta(p^0_i)\,\hat \delta(p^2_i-m^2_i),\,\,\,\,\, D=4-2\epsilon,$$
for some energy scale $\mu$.
We also set $c = 1,\,\,\,\hbar=1,\,\,\,\epsilon^{0123}=1$, and work in Heaviside–Lorentz units for the electric charges. Fourier transforms are defined by 
$$
    f(x)=\int\! \hat\dd^Dk\ e^{-ik\cdot x} \tilde{f}(k),\, \qquad \tilde{f}(k)=\int\! \dd^Dx\ e^{ik\cdot x}f(x),
$$
where, unless specified, limits of the integrals are $(-\infty, \infty)^{D}$.   We
use the notation $a^\mu$ to indicate the particle acceleration while $\sa^\mu$ denotes its spin (pseudo) vector.
Finally, we do not divide by $k!$  when symmetrising or antisymmetrising over $k$ indices. For instance when $k=2$ we would write $x^{(\mu}y^{\nu)}=x^{\mu}y^{\nu}+y^{\mu}x^{\nu}$ and similarly $x^{[\mu}y^{\nu]}=x^{\mu}y^{\nu}-y^{\mu}x^{\nu}$.

\section{General framework}\label{general}

In this work, we derive spinning radiation reaction equations by regularising the classical self-field and independently test the resulting dynamics using Compton cuts. This section establishes the classical
ingredients shared by these calculations. We first review the spinless self-field problem and
the local expansion used to extract its finite part. We then specify the $\sqrt{\text{Kerr}}$ worldline
dynamics and electromagnetic current through quadratic order in spin. Finally, we construct
a waveform kernel that is linear in the worldline deflections and can be evaluated using
either the leading conservative motion or its radiation-reaction correction.

\subsection{Radiation and reaction}\label{ssrr}
We first consider a structureless point charge, with worldline $z^\mu(\tau)$, parametrised by proper time. We write
$u^\mu=\dot z^\mu$ and $a^\mu=\dot u^\mu$, with $u^2=1$. The charge
sources the current
\[
 J^\mu(x)
    =Q\int \dd\tau\,u^\mu(\tau)
      \delta^{(4)}\bigl(x-z(\tau)\bigr).
    \label{eq:spinless-current}
\]
This current sources a field strength $ F_{\mu\nu}=\partial_\mu A_\nu-\partial_\nu A_\mu$. The Maxwell equation $\partial_\mu F^{\mu\nu}=J^\nu $ reduces in Lorenz gauge, $\partial_\mu A^\mu=0$, to the wave equation
$\partial^2 A^\mu=J^\mu$. The retarded solution is
\[
 A_{\mathrm{ret}}^\mu(x)
    =\int \dd^4x'\,G_{\mathrm{ret}}(x-x')J^\mu(x')
    =Q\int \dd \tau\,G_{\mathrm{ret}}\bigl(x-z(\tau)\bigr)
                    u^\mu(\tau)
     .
    \label{eq:retarded-self-field}
\]
Here $G_{\rm ret}$ is the retarded Green's function. It is strictly supported on the future lightcone of its source and is given by
\[
G_{\rm ret}(x-x')= \frac{\Theta\left(t-t^{\prime}\right)}{2 \pi} \delta\left(\left(x-x^{\prime}\right)^2\right)=\frac{\delta\left(t-t^{\prime}-\left|\mathbf{x}-\mathbf{x}^{\prime}\right|\right)}{4 \pi\left|\mathbf{x}-\mathbf{x}^{\prime}\right|}.
\]
The convolution \eqref{eq:retarded-self-field} of the Green's function with the source produces a field $ A_{\mathrm{ret}}^\mu(x)$ that is singular on the worldline $z(\tau)$. This divergence becomes problematic when the charge is allowed to respond to its own field, where the singularity at $x = z(\tau)$ renders the Lorentz force equation,
\[
m a^\mu  = Q F^{\mu \nu}_{\rm ret}(z(\tau)) u_\nu(\tau),
\]
ill-defined \cite{Poisson:2011nh,Dirac:1938nz}. However, it is precisely this limit that encodes the physics of radiation reaction: the motion of a charge induced by its own radiated field. The singularity at $x = z(\tau)$ can be cured by 
introducing the advanced field $F_{\mathrm{adv}}^{\mu\nu}$, supported on the past lightcone of its source via the advanced Green's function $G_{\mathrm{adv}}(x-x')
=\Theta(t'-t)\delta((x-x')^2)/(2\pi)$. The key insight is that $F_{\mathrm{adv}}^{\mu\nu}$ suffers from exactly the same singularity as $F_{\mathrm{ret}}^{\mu\nu}$ in the limit $x \rightarrow z(\tau)$. In particular, the \emph{regular} field $F_{\mathrm R}^{\mu\nu}$ defined by the difference of the retarded and advanced fields is finite in this limit and encodes the physical effects of radiation reaction. This observation inspires the decomposition
\[\label{eq:DiracDeomposition}
 F_{\mathrm{ret}}^{\mu\nu} =F_{\mathrm S}^{\mu\nu}+F_{\mathrm R}^{\mu\nu},
\]
where $F_{\mathrm R}^{\mu\nu}$ and $F_{\mathrm S}^{\mu\nu}$ are the regular and singular parts:
\[
   F_{\mathrm R}^{\mu\nu} =\frac12\bigl(F_{\mathrm{ret}}^{\mu\nu} -F_{\mathrm{adv}}^{\mu\nu}\bigr), \quad F_{\mathrm S}^{\mu\nu}=\frac12\bigl(F_{\mathrm{ret}}^{\mu\nu} +F_{\mathrm{adv}}^{\mu\nu}\bigr).
\]
The singular field $F_{\mathrm S}^{\mu\nu}$ diverges when $x = z(\tau)$ and is typically absorbed into a 
renormalisation of the mass $m$ or other parameters. The physical equation encoding radiation reaction is induced by the regular field $F_{\mathrm R}^{\mu\nu}$ and is given by
\[
m a^\mu = Q F_R^{\mu \nu}(z(\tau)) u_\nu.
\]
The problem of determining the correct radiation reaction equation therefore boils down to calculating the finite quantity
\[
F_{\mathrm R}^{\mu\nu}(z(\tau))&=\lim_{x\rightarrow z(\tau)} \frac12\bigl(F_{\mathrm{ret}}^{\mu\nu}(x)
               -F_{\mathrm{adv}}^{\mu\nu}(x)\bigr), \\
&= \lim_{x\rightarrow z(\tau)} \frac{Q}{2}\int \dd\tau^\prime\, \partial ^{[\mu}\Bigl(G_{\mathrm{ ret}}\bigl(x-z(\tau')\bigr)- G_{\mathrm{ adv}}\bigl(x-z(\tau')\bigr)\Bigr) u^{\nu]}(\tau').               
\]
As we will see in the next section, the coincidence limit described above will end up yielding a series of cut integrals that closely resemble unitarity cuts.
Applying this subtraction procedure (see eg. \cite{Poisson:1999tv}) then leads to the regular field strength
 \[
 F_{\mathrm R}^{\mu\nu}\bigl(z(\tau)\bigr)
    =\frac{Q}{6\pi}
  \bigl(\dot a^\mu u^\nu-\dot a^\nu u^\mu\bigr),
     \label{eq:spinless-regular-field}
 \]
which gives rise to the famous ALD equation
\[ 
 m a^\mu
    =\frac{Q^2}{6\pi}
      \bigl(\dot a^\mu+a^2u^\mu\bigr).
    \label{eq:ald-equation}
\]
It is straightforward to see that the decomposition \eqref{eq:DiracDeomposition} isolates the dissipative content of the field in $ F_{\mathrm R}^{\mu\nu}\bigl(z(\tau)\bigr)$. To see this at the level of the potential, write
\[
A_\zeta^\mu(t,x) \sim \int d^3 x'\,  \frac{J^\mu(t-\zeta | \Vec{x} - \Vec{x}'|)}{ R} , \quad R = | \Vec{x} - \Vec{x}'|,
\]
where $\zeta = \pm 1$ for retarded and advanced fields respectively. Expanding this expression in $R/T \ll 1$, where $T$ is the source variation timescale, separates terms even and odd in $\zeta$. Only the odd terms in $\zeta$ survive in the combination $A_{\rm ret}^\mu(t,x) - A_{\rm adv}^\mu(t,x) $, carrying the time-asymmetric physics of dissipation, see eg. \cite{Poisson:1999tv, Detweiler:2002mi} for a detailed derivation and also \cite{Poisson:2011nh,Whiting:2003tx} for applications to curved space. The regular field $F_{\mathrm R}^{\mu\nu}$ therefore cleanly isolates the dissipative content of the Lorentz force.

Rather than subtracting the advanced solution, we will extract the same finite self-field using a local expansion of the retarded solution, following a technique by Coleman \cite{RM-2820-PR} and its momentum-space implementation in
\cite{Elkhidir:2023dco}. 

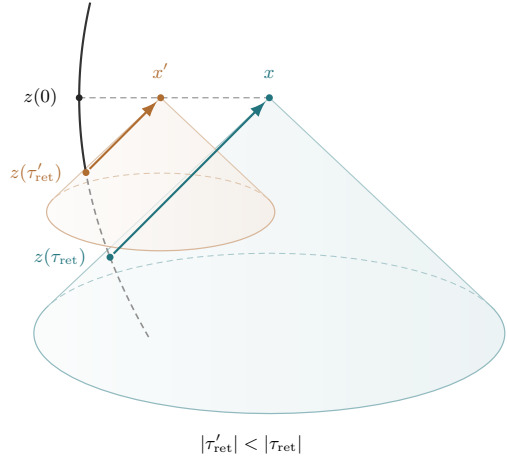
\begin{figure}[t]
    \centering
    \resizebox{0.45\linewidth}{!}{%
        \begin{tikzpicture}[x=1.55cm,y=1.55cm,>=Latex,font=\small]
  \definecolor{farcone}{RGB}{34,117,126}
  \definecolor{nearcone}{RGB}{176,109,50}
  \def\kappa{0.22}
  \def\radius{2.1}
  \def\nearradius{0.9}
  \def\flattening{0.34}
  \pgfmathsetmacro{\delay}{(sqrt(1+2*\kappa*\radius)-1)/\kappa}
  \pgfmathsetmacro{\neardelay}{(sqrt(1+2*\kappa*\nearradius)-1)/\kappa}
  \pgfmathsetmacro{\sourcexi}{\radius-\delay}
  \pgfmathsetmacro{\nearsourcexi}{\nearradius-\neardelay}
  \pgfmathsetmacro{\angle}{asin(\flattening)}
  \pgfmathsetmacro{\leftangle}{180-\angle}
  \pgfmathsetmacro{\rightangle}{360+\angle}
  \coordinate (O) at (0,0);
  \coordinate (X) at (\radius,0);
  \coordinate (Xp) at (\nearradius,0);
  \coordinate (R) at (\sourcexi,-\delay);
  \coordinate (Rp) at (\nearsourcexi,-\neardelay);

  \newcommand{\pastcone}[3]{%
    \begin{scope}
      \pgfmathsetmacro{\height}{\flattening*(#2)}
      \coordinate (apex) at (#1,0);
      \coordinate (left) at ({#1+(#2)*cos(\leftangle)},
                              {-(#2)+\height*sin(\leftangle)});
      \coordinate (right) at ({#1+(#2)*cos(\angle)},
                               {-(#2)+\height*sin(\angle)});
      \shade[left color=#3!2,right color=#3!12,opacity=.7]
        (apex)--(left)
        arc[start angle=\leftangle,end angle=\rightangle,
            x radius=#2,y radius=\height]--cycle;
      \draw[#3!40,line width=.55pt] (apex)--(left);
      \draw[#3!40,line width=.55pt] (apex)--(right);
      \draw[#3!40,line width=.5pt,densely dashed]
        (right) arc[start angle=\angle,end angle=\leftangle,
                    x radius=#2,y radius=\height];
      \draw[#3!50,line width=.6pt]
        (left) arc[start angle=\leftangle,end angle=\rightangle,
                   x radius=#2,y radius=\height];
    \end{scope}}

  \pastcone{\radius}{2.6}{farcone}
  \pastcone{\nearradius}{1.26}{nearcone}

  \draw[black!43,line width=.8pt,densely dashed,
        domain=-2.64:-\neardelay,samples=70,variable=\s]
    plot ({.5*\kappa*\s*\s},{\s});
  \draw[black!80,line width=1.05pt,
        domain=-\neardelay:1.05,samples=65,variable=\s]
    plot ({.5*\kappa*\s*\s},{\s});

  \draw[black!40,line width=.6pt,densely dashed] (O)--(X);
  \draw[farcone,line width=1.05pt,->,
        shorten <=2.5pt,shorten >=2.5pt] (R)--(X);
  \draw[nearcone,line width=1.05pt,->,
        shorten <=2.5pt,shorten >=2.5pt] (Rp)--(Xp);

  \fill[black!85] (O) circle (1.65pt);
  \fill[farcone] (R) circle (1.7pt);
  \fill[farcone] (X) circle (1.7pt);
  \fill[nearcone] (Rp) circle (1.7pt);
  \fill[nearcone] (Xp) circle (1.7pt);
  \node[left=7pt] at (O) {$z(0)$};
  \node[left=8pt,text=farcone] at (R) {$z(\tau_{\rm ret})$};
  \node[left=8pt,text=nearcone] at (Rp) {$z(\tau'_{\rm ret})$};
  \node[above=6pt,text=farcone] at (X) {$x$};
  \node[above=6pt,text=nearcone] at (Xp) {$x'$};
  \node at (1.9,-3.82) {$|\tau'_{\rm ret}|<|\tau_{\rm ret}|$};
\end{tikzpicture}
    }
    \caption{Past lightcones of the observation points $x$ and $x'$.
    As the observation point approaches the worldline, the retarded
    source point approaches $z(0)$, so that
    $|\tau'_{\rm ret}|<|\tau_{\rm ret}|$.}
    \label{fig:retarded-limit}
\end{figure}

To motivate this, consider a point $z(0)$  on the worldline, placed on the origin for convenience. Let $x$ be the observation point at a perpendicular distance $r$ from $z(0)$. We are interested in taking the limit $x \rightarrow z(0)$ of the retarded field strength sourced by this worldline, ie the small $r$ limit. The retarded field at $x$ depends on the worldline point at the retarded time $\tau_{\rm ret}$, so that $F_{\rm ret}^{\mu \nu}$ is a function of $z^\mu(\tau_{\rm ret})$. 
The lightcone condition $(x-z(\tau_{\rm ret}))^2=0$ imposes $\tau_{\rm ret} \sim r$. As such, we can identify the small $r$ and small $\tau_{\rm ret}$ limits, see figure~\ref{fig:retarded-limit}. We can therefore expand worldline quantities in a small-$\tau$ expansion 
\[
z^\mu(\tau_{\rm ret})
    =z^\mu(0)+u^\mu\tau_{\rm ret}
     +\frac12a^\mu\tau_{\rm ret}^2
     +\frac16\dot a^\mu\tau_{\rm ret}^3+\cdots,
    \label{eq:local-worldline-expansion}
\]
and consequently expand the retarded field $F_{\rm ret}^{\mu \nu}$ in small $\tau_{\rm ret}$. This is most conveniently carried out in momentum space, where one can exploit dimensional regularisation to write
\[
F_{\rm ret}^{\mu \nu}(z(0))=i Q \int \hat{\dd}^D k \int_{-\infty}^0 \mathrm{~d} \tau\, e^{i k \cdot z(\tau)} \frac{k^{[\mu} u^{\nu]}(\tau)}{\left(k^0+i \epsilon\right)^2-\mathbf{k}^2},
\]
The limit $r \rightarrow 0$ is therefore repackaged as an expansion in $\tau_{\rm ret} \sim k^{-1} \ll 1$, where all worldline quantities are expanded in small $\tau_{\rm ret}$. Instead of subtracting the advanced contribution $F_{\rm adv}^{\mu \nu}$, our divergences are now regulated by the dimensional regulator $D=4-2\epsilon$. This has the advantage of automatically discarding scaleless integrals, which form the bulk of the integrals emerging from the small $\tau_{\rm ret}$ expansion above. It is then straightforward to show (see \cite{Elkhidir:2023dco}) that evaluating the non-vanishing scalar integrals leads precisely to the renormalised field strength \eqref{eq:spinless-regular-field} responsible for the ALD equation. In the following sections, we use this renormalisation procedure to derive new radiation reaction equations for spinning bodies. First, we give a brief overview of the $\sqrt{\rm Kerr}$ theory.

\subsection{The $\sqrt{\rm Kerr}$ theory}

We describe the spinning charge by a worldline action, retaining terms through quadratic order in spin. Specifically, we work with the $\sqrt{\rm Kerr}$ \cite{Monteiro:2014cda, Arkani-Hamed:2019ymq, Menezes:2022tcs, Scheopner:2023rzp,Kim:2024grz, deFirmian:2026mln}  theory, defined by the action
\[\label{eq:SpinningAction}
\mathcal S&=-\frac14\int \dd^4x\,F_{\mu\nu}F^{\mu\nu}+\mathcal S_{\rm wl},\\
\mathcal S_{\rm wl}&=\int \dd\lambda\,
\left[-p_\mu z'{}^\mu-QA_\mu z'^\mu+L_{\rm spin}
+\frac{\alpha}{2}\big(p^2-\mathcal{M}^2\big)+\beta\,p\cdot \sS\right],
\]
where primes denote derivatives with respect to the worldline parameter $\lambda$. $z^\mu(\lambda)$ is the worldline trajectory, $A^\mu(x)$ is the gauge field and $F_{\mu \nu}=\partial_{[\mu} A_{\nu]}$ is the field strength tensor and $\sS$ is the spin of the massive body. We will often take the worldline parameter $\lambda$ to coincide with the proper time $\tau$  and   take $\alpha=1/\mathcal{M}$. In this case, we will denote derivatives with respect to $\tau$ using overdots and further define
\[
u^\mu = \dot{z}^\mu, \quad a^\mu = \dot{u}^\mu,
\]
for the four-velocity and four-acceleration respectively. $L_{\rm spin}$ is the spin kinetic term defined by
\[
L_{\text {spin }}=\frac{1}{2} \epsilon_{\mu \nu \rho \sigma} \ph^\mu \sS^\nu \Omega^{\rho \sigma}, \quad \Omega^{\mu \nu}=\eta^{A B} \Lambda^\mu{ }_A \frac{\dd\Lambda^\nu{ }_B}{\dd \lambda},
\]
where $\Lambda$ is an orthonormal tetrad attached to the body $\eta_{\mu \nu} \Lambda^\mu{ }_A \Lambda^\nu{ }_B=\eta_{A B} $. The Lagrange multipliers $\alpha$ and $\beta$ enforce the on-shell condition and the Tulczyjew–Dixon spin-supplementary condition:
\[
p^2=\mathcal{M}^2,\qquad p\cdot \sS=0,\qquad \ph^\mu\equiv\frac{p^\mu}{\sqrt{p^2}}=\frac{p^\mu}{\mathcal{M}}.
\]
where $\mathcal{M}$ is the dynamical mass function. Specifically, $\mathcal{M}^2$ is a function of the worldline variables given by
\[\label{eq:DynamicalMass}
&\mathcal{M}^2=m^2+2 Q^{\star} F_{\mu \nu} \sS^\mu \ph^\nu-\frac{Q}{m} \sS^\rho \partial_\rho F_{\mu \nu} \sS^\mu \ph^\nu+O\left(\sS^3\right),\\
\]
for the $\sqrt{\rm Kerr}$ theory, where
\[
{}^\star\!F_{\mu\nu}  =\frac{1}{2}\epsilon_{\mu \nu \rho \sigma}F^{\rho \sigma},
\]
is the dual field strength tensor. We work with the convention where $\epsilon^{0123}=1$ (so that $\epsilon_{0123}=-1)$. Here the constant $m$ is the mass of the particle in the absence of spin couplings. The linear and quadratic in spin terms describe the dipole and quadrupole couplings, while the monopole coupling is captured by the standard term $QA_\mu z'^\mu$ in \eqref{eq:SpinningAction}. 

It is important to distinguish the field expansion of $\mathcal{M}^2$ from that of $\mathcal{M}$. Specifically, for the $\sqrt{\rm Kerr}$ theory $\mathcal{M}^2$ is linear in $F_{\mu \nu}$ through quadratic order in spin\footnote{The absence of $F^2$ terms in $\mathcal{M}^2$ to this order is fixed by a matching calculation to the $\sqrt{\rm Kerr}$ Compton amplitude, see eg. \cite{Scheopner:2023rzp}.}.  $\mathcal{M}$, however, acquires an $\mathcal{O}(F^2)$ term through the expansion of the square-root $\sqrt{\mathcal{M}^2}$ to quadratic order in spin:
\[
\mathcal{M}=m+\frac{Q}{m}{ }^{\star} F_{\mu \nu} \sS^\mu \ph^\nu-\frac{Q}{2 m^2} \sS^\rho \partial_\rho F_{\mu \nu} \sS^\mu \ph^\nu-\frac{Q^2}{2 m^3}\left({ }^{\star} F_{\mu \nu} \sS^\mu \ph^\nu\right)^2+\mathcal O\left(\sS^3\right) .
\]
As we will see below, the $\mathcal{O}(F^2\sS^2)$ term in $\mathcal{M}$ contributes to the conserved current at linear order in $F$, and is therefore retained in our calculations. The mass-shell conditions further imply the identities
\[
p^2=m^2+2m(\mathcal{M}-m)+\mathcal{O}(F^2),\qquad
p\cdot\dot p=\mathcal{M}\mathcal{ \dot M},\qquad \ph\cdot\dot p= \mathcal{\dot M}.
\label{eq:mass_evolution}
\]
\paragraph{Four-velocity versus momentum.}
A consequence of the dynamical mass function is that the unit momentum $\ph^\mu = p^\mu/\mathcal{M}$ and the four-velocity $u^\mu = \dd z^\mu/\dd\tau$ cease to be parallel. This is most intuitively understood by considering a spinning body as an extended object. The motion of such an object consists of the motion of its center-of-mass (fixed by the SSC), and internal motion about the center of momentum. Since $z^\mu$ is defined by the center-of-mass (COM), $u^\mu$ is necessarily tangent to the motion of the COM. The momentum $p^\mu$ (and therefore $\ph^\mu$) is also sensitive to the internal motion of the body, and need not be parallel to $u^\mu$. 

\paragraph{Equations of motion}
The aforementioned distinction between $\ph^\mu$ and $u^\mu$ becomes manifest when computing the equations of motion. To this end, we vary the action \eqref{eq:SpinningAction} with respect to $z$, $p$ and $\sS$, with the field held fixed. This gives, at linear order in $F$,
\[\label{eq:TreeEOM}
\dot p^\mu&=Q\,F^\mu{}_\nu u^\nu+\partial^\mu \mathcal{M},\\
\dot \sS^\mu&=-\ph^\mu\,(\dot{\ph}\cdot \sS)-\epsilon^{\mu\nu\rho\sigma}\ph_\nu \sS_\rho\,\frac{\partial \mathcal{M}}{\partial \sS^\sigma},\\
u^\mu-\ph^\mu&=-\frac{1}{\mathcal{M}}P^{\mu\nu}\frac{\partial \mathcal{M}}{\partial \ph^\nu}+\frac{1}{\mathcal{M}}\sS^\mu \ph^\nu\frac{\partial \mathcal{M}}{\partial \sS^\nu}
-\frac{1}{\mathcal{M}^2}\epsilon^{\mu\nu\rho\sigma}\ph_\nu \sS_\rho\,\partial_\sigma \mathcal{M}-\frac{Q}{\mathcal{M}^2}P^\mu{}_\nu\,{}^\star\!F^{\nu\rho}\sS_\rho ,
\]
with $p^\mu=\mathcal{M}\ph^\mu$, $ \mathcal{\dot M}=\ph\cdot\dot p$ and $P^{\mu\nu}\equiv\eta^{\mu\nu}-\ph^\mu \ph^\nu$ the projector orthogonal to $\ph^\mu$. The first term of the spin equation enforces the SSC. The derivatives of the dynamical mass function in~(\ref{eq:DynamicalMass}) are
\[\label{eq:DynamicalMassDerivatives}
\frac{\partial \mathcal{M}}{\partial \ph^\nu}&=\frac Qm\,{}^\star\!F_{\mu\nu}\sS^\mu-\frac{Q}{2m^2}\,\sS^\rho\partial_\rho F_{\mu\nu}\sS^\mu,\\
\frac{\partial \mathcal{M}}{\partial \sS^\sigma}&=\frac Qm\,{}^\star\!F_{\sigma\nu}\ph^\nu-\frac{Q}{2m^2}\Big(\partial_\sigma F_{\mu\nu}\sS^\mu \ph^\nu+\sS^\rho\partial_\rho F_{\sigma\nu}\ph^\nu\Big),\\
\partial_\sigma \mathcal{M} &=\frac Qm\,\partial_\sigma{}^\star\!F_{\mu\nu}\sS^\mu \ph^\nu-\frac{Q}{2m^2}\,\sS^\rho\partial_\rho\partial_\sigma F_{\mu\nu}\sS^\mu \ph^\nu.
\]
These equations are the electromagnetic analogue of the MPD equations in gravity \cite{Mathisson:1937zz,Dixon:1970zza, Papapetrou:1951pa, Scheopner:2023rzp}. Their truncation to linear order in spin is also known as the BMT equation \cite{Bargmann:1959gz,Jackson}.

\paragraph{Current}
The worldline action also defines the conserved current through the relation
\[
J^\mu=-\delta\mathcal S_{\rm wl}/\delta A_\mu, \quad \partial_\mu F^{\mu\nu}=J^\nu.
\]
Explicit variation of the dynamical mass function yields
\[
\delta_A \mathcal{M}= & \frac{Q}{m}\left(1-\frac{Q}{m^2}{ }^{\star} F_{\gamma \delta} \sS^\gamma \ph^\delta\right) \sS^\alpha \ph^\beta \delta_A{ }^{\star} F_{\alpha \beta}  -\frac{Q}{2 m^2} \sS^\rho \sS^\alpha \ph^\beta \partial_\rho \delta_A F_{\alpha \beta}+O\left(\sS^3\right) ,
\]
so that the current through linear order in $F$ is given by
\[\label{current}
J^\mu(x)=Q\int \dd\tau\Big[u^\mu\,\delta^4(x-z)+\left( 1- \frac{Q}{m^2}{ }^{\star} F_{\alpha \beta} \sS^\alpha \ph^\beta\right)D^{\mu\nu}\partial_\nu\delta^4(x-z)+C^{\mu\nu\rho}\partial_\nu\partial_\rho\delta^4(x-z)\Big],
\]
where the respective dipole and quadrupole terms are
\[
D^{\mu\nu}=-\frac1m\,\epsilon^{\mu\nu\alpha\beta}\ph_\alpha \sS_\beta,\qquad
C^{\mu\nu\rho}=\frac{1}{2m^2}\,\sS^\rho\left(\sS^\mu \ph^\nu-\sS^\nu \ph^\mu\right).
\]
This current takes an especially simple form upon recognizing the exponential structure of the $\sqrt{\rm Kerr}$ theory. This is most easily seen in momentum space where 
\[
\tilde J^\mu(k)=\int \dd^4x\,e^{ik\cdot x}J^\mu(x).
\]
Defining 
\[
 (\sa * \partial)^\mu{ }_\nu \equiv-\epsilon^{\mu \alpha \rho \sigma} \eta_{\alpha \nu} \sa_\rho \partial_\sigma, \quad \mathsf a^\mu = \mathsf S^\mu/m,
\]
it is straightforward to check that
\[\label{mtrxexponential}
 \exp(\sa * \partial)^\mu_\nu \ph^\nu \delta^4(x-z)= \left(\ph^\mu + D^{\mu\nu}\partial_\nu+C^{\mu\nu\rho}\partial_\nu\partial_\rho \right) \delta^4(x-z) + \mathcal{O}(\sa^3),
\]
up to off-shell terms that are not relevant for our purposes.
The on-shell current therefore takes the form
\[\label{eq:Root-Kerr-Current}
J^\mu(x)=Q\int \dd \tau\left[ (u-\ph)^\mu -\frac{Q}{m}{ }^{\star} F_{\alpha \beta} \sa^\alpha \ph^\beta D^{\mu\nu}\partial_\nu+ \exp(\sa * \partial)^\mu_\nu \ph^\nu  \right]\delta^4(x-z).
\]
When evaluated at leading order, we can set $\ph^\mu = u^\mu$ and discard $\mathcal{O}(F)$ terms to arrive at
\[
J^\mu(x)_{\rm LO}=Q\int \dd\tau  \, \exp(\sa * \partial)^\mu_\nu u^\nu\delta^4(x-z),
\]
leading to the well-known exponentiation of the leading $\sqrt{\text{Kerr}}$ current/three-point amplitude \cite{Arkani-Hamed:2019ymq, Vines:2017hyw,Menezes:2022tcs}. For our purposes, our calculations require perturbing the worldline variables up to $\mathcal{O}(\sa^2)$ where the distinction between $\ph^\mu$ and $u^\mu$ appears. We therefore use the form \eqref{eq:Root-Kerr-Current} for our calculations. 

\subsection{Spinning waveforms}\label{sub:swaves}

Having specified our dynamics, we are now in a position to calculate the waveform sourced by a spinning particle scattering against a scalar (non-spinning) particle. We will take the background particle to be particle 1 and the spinning and radiating particle to be particle 2.
We will specifically calculate the waveshape defined by
\[
\alpha_\eta(k)=-i\,\varepsilon^{*}_\eta(k)\cdot\tilde J(k),\qquad k^2=0,
\]
with the current sourced by the spinning particle 2:
\[
\tilde J_2^\mu(k)=Q_2\!\int\! \dd\tau\,e^{ik\cdot z_2(\tau)}\Big[u_2^\mu+\frac{m_2}{\mathcal{M}_2}\,i\,\epsilon^\mu(k,\ph_2,\mathsf a_2)
+\tfrac12(\mathsf a_2\cdot k)\big((\mathsf a_2\cdot k)\ph_2^\mu-(\ph_2\cdot k)\mathsf a_2^\mu\big)\Big].
\label{eq:Jk}
\]
where $m_2/ \mathcal{M}_2=1-\frac{Q_2}{m_2^2}{}^\star F_{\alpha\beta}\sS_2^\alpha \ph_2^\beta$ to the order required. 
Note that in the above equation (and in what follows) we have introduced the convenient notation
\[
\epsilon(a,b,c,d)=\epsilon_{\alpha\beta\gamma\delta}a^\alpha b^\beta c^\gamma d^\delta, \,\,\, \epsilon_\mu(a,b,c)= \epsilon_{\mu\alpha\beta\gamma}a^\alpha b^\beta c^\gamma ,\,\,\,\epsilon_{\mu\nu}(a,b)= \epsilon_{\mu\nu\alpha\beta}a^\alpha b^\beta.
\] 
The calculation of the waveform simply amounts to expanding this current to the required order in $Q$ and $\sa$, taking into account the deflections $\delta z_2,\delta u_2,\delta \ph_2$ and $\delta \sa_2$ sourced by the background field, which in our case is the Coulombic field sourced by particle 1. At leading order, the current is evaluated by substituting the undeflected trajectory $z_2=b_2+u_2\tau$, along which $u_2^\mu=\ph_2^\mu$ is constant (in what follows $u_2$ and $u_1$ without arguments always denote these constant background velocities), yielding
\[
\tilde J_2^\mu(k)=Q_2\,e^{ik\cdot b_2}\hat{\delta}(k\cdot u_2)\,j_2^\mu,\qquad j_2^\mu\equiv u_2^\mu+i\,\epsilon^\mu(k,u_2,\mathsf a_2)+\tfrac12(\mathsf a_2\cdot k)\big((\mathsf a_2\cdot k)u_2^\mu-(u_2\cdot k)\mathsf a_2^\mu\big),
\]
where $j_2^\mu$ is the $\mathcal O(\mathsf a_2^2)$ truncation of $\big[\exp(-i\,\mathsf a_2*k)\big]^\mu{}_\nu u_2^\nu$. 
Beyond this, our task is to calculate the aforementioned deflections to the required order. In this section we calculate the deflections at leading order in $Q$ and to quadratic order in spin. The deflections at subleading order require knowledge of the radiation reaction equations which we derive in the following sections. 

Since most of our calculation is agnostic to the nature of the background field, we will find it useful to work with a generic field strength $F_{\mu \nu}$ in intermediate steps. Moreover, since the waveform is linear in this background field, we can exploit this linearity (and the linearity of electromagnetic interactions) to write our background field as a single plane wave 
\[
\mathsf A^\mu(x)=\mathsf{\tilde A}^\mu(q) e^{-iq\cdot x},\qquad F_{\mu\nu}(x)=f_{\mu\nu}(q)\,e^{-iq\cdot x},\qquad
f_{\mu\nu}=-i\big(q_\mu\mathsf{\tilde A}_\nu-q_\nu \mathsf{\tilde A}_\mu\big).
\]
Since any (reasonable) background field can be constructed by a Fourier transform, we can specify our background at the end of the calculation by summing over these plane wave modes
\[
F_{\mu \nu}(x) = \int \hd^4 q\, f_{\mu \nu}(q) \,e^{- i q \cdot x}.
\]
As we will see, this will considerably simplify our task. In particular, using substitutions of the form 
$\partial_\rho F_{\mu\nu}\to-iq_\rho f_{\mu\nu}$ in the equations of motion turns them into purely algebraic equations which can be readily solved. We take the leading trajectory to be $z_2 = b_2 + u_2 \tau$ so that at leading order
\[
F_{\mu \nu} = f_{\mu\nu}e^{-iq\cdot b_2}e^{-i\omega\tau}, \quad \omega\equiv q\cdot u_2.
\]
At this order, we can therefore apply the substitutions
\[
\frac{\dd}{\dd\tau}\;\longrightarrow\;-i\omega,\qquad \int^\tau \dd\tau'\;\longrightarrow\;\frac{i}{\omega +i \epsilon},
\label{eq:dinttau}
\]
inside the equations of motion.
We will also find it convenient to introduce the projector 
\[
P_2^{\mu\nu}\equiv\eta^{\mu\nu}-u_2^\mu u_2^\nu,
\]
and the notation
\[
&f(\mathsf a_2,u_2)\equiv f_{\mu\nu}\mathsf a_2^\mu u_2^\nu,\qquad &&{}^\star\! f(\mathsf a_2,u_2)\equiv{}^\star\! f_{\mu\nu}\mathsf a_2^\mu u_2^\nu,\\
&(f\!\cdot\! u_2)^\mu\equiv f^{\mu}{}_{\nu}u_2^\nu,\qquad &&(\mathsf a_2\!\cdot\! f)_\nu\equiv \mathsf a_2^\mu f_{\mu\nu} .
\]
We now express our equations of motion for our single Fourier mode. These will then determine the deflections necessary for evaluating the current in \eqref{eq:Jk}.

First, we use \eqref{eq:DynamicalMass} to evaluate the dynamical mass function on the worldline. Writing $\mathcal{M}_2=m_2+\delta \mathcal{M}_2$ we have
\[
\delta \mathcal{M}_2= Q_2\,{}^\star\! f(\mathsf a_2,u_2)+\frac{iQ_2}{2}\,(q\cdot\mathsf a_2)\,f(\mathsf a_2,u_2).
\label{eq:dM}
\]
Turning to the momentum, with $p_2=\mathcal{M}_2 \ph_2$ and $\dot{ \mathcal{M}}_2=\ph_2\cdot\dot p_2$, the first line of \eqref{eq:TreeEOM}, combined with the substitution in \eqref{eq:dinttau} gives
\[\label{eq:du}
\;\delta \ph_2^\mu=\frac{i}{m_2\omega}\Big[Q_2\,(f\!\cdot\! u_2)^\mu-i\,(q^\mu-\omega u_2^\mu)\,\delta \mathcal{M}_2\Big].\;
\]
Applying similar substitutions to the second equation in \eqref{eq:TreeEOM} combined with \eqref{eq:DynamicalMassDerivatives} yields the following spin deflection
\[\label{eq:da}
\delta\mathsf a_2^\mu=-u_2^\mu\,(\mathsf a_2\cdot\delta \ph_2)-\frac{iQ_2}{m_2\omega}\,\epsilon^{\mu}(u_2,\mathsf a_2,X),\qquad
X_\sigma={}^\star\! f_{\sigma\nu}u_2^\nu+\frac i2\Big[q_\sigma\,f(\mathsf a_2,u_2)+(q\cdot\mathsf a_2)\,f_{\sigma\nu}u_2^\nu\Big].
\]
Finally, we calculate the $\mathcal{O}(\sa_2^2)$ correction $u^\mu - \ph^\mu$ from the third line of \eqref{eq:TreeEOM} which gives
\[\label{eq:hidden}
(u_2-\ph_2)^\mu=\frac{iQ_2}{2m_2}\Big[\omega\,\mathsf a_2^\mu\,f(\mathsf a_2,u_2)-(q\cdot\mathsf a_2)\,P_2^{\mu\nu}(\mathsf a_2\!\cdot\! f)_\nu
+2\,\epsilon^\mu(u_2,\mathsf a_2,q)\;{}^\star\! f(\mathsf a_2,u_2)\Big].
\]
The above equations suffice to also determine the velocity and position through
\[
\delta u_2=\delta \ph_2+(u_2-\ph_2),\qquad \delta z_2=\frac{i}{\omega}\,\delta u_2 .
\label{eq:dz}
\]
We can now proceed to substituting the leading order deflections in the current \eqref{eq:Jk}. Specifically, substituting
\[
z_2=b_2+u_2\tau+\delta z_2,\quad u_2\to u_2+\delta u_2,\quad \ph_2= u_2+\delta \ph_2,\quad
\mathsf a_2\to\mathsf a_2+\delta\mathsf a_2,\quad \mathcal{M}_2=m_2+\delta \mathcal{M}_2,
\]
expanding $e^{ik\cdot z_2}=e^{ik\cdot(b_2+u_2\tau)}(1+ik\cdot\delta z_2)$, and truncating at linear order in the deflections we obtain the leading order correction to the waveshape
\[
-i\,\varepsilon^*\!\cdot\tilde J_2^{(1)}(k)=-i\,e^{i(k-q)\cdot b_2}\,\hat\delta\big(u_2\cdot(k-q)\big)\;\mathcal W,
\]
where the kernel $\mathcal{W}$ is given by
\[\label{eq:WaveformDeflection}
&\mathcal W[\delta u_2,\delta z_2,\delta \ph_2,\delta\mathsf a_2,\delta \mathcal{M}_2]=Q_2\Big[\;\varepsilon^*\!\cdot\delta u_2
\;+\;i\,(k\cdot\delta z_2)\;\varepsilon^*\!\cdot j_2\\
&+\;i\,\epsilon(\varepsilon^*,k,\delta \ph_2,\mathsf a_2)+\tfrac12(\mathsf a_2\cdot k)\big[(\mathsf a_2\cdot k)\,\varepsilon^*\!\cdot\delta \ph_2-(\delta \ph_2\cdot k)\,\varepsilon^*\!\cdot\mathsf a_2\big]\\
&+\;i\,\epsilon(\varepsilon^*,k,u_2,\delta\mathsf a_2)+\tfrac12(\delta\mathsf a_2\cdot k)\big[(\mathsf a_2\cdot k)\,\varepsilon^*\!\cdot u_2-(u_2\cdot k)\,\varepsilon^*\!\cdot\mathsf a_2\big]\\
&+\tfrac12(\mathsf a_2\cdot k)\big[(\delta\mathsf a_2\cdot k)\,\varepsilon^*\!\cdot u_2-(u_2\cdot k)\,\varepsilon^*\!\cdot\delta\mathsf a_2\big]-\;\frac{\delta \mathcal{M}_2}{m_2}\;i\,\epsilon(\varepsilon^*,k,u_2,\mathsf a_2)\Big].
\]
So far, the expressions above determine the tree-level current regardless of the background field $F_{\mu \nu}$. Now, we specialise to the background sourced by particle 1, which is taken to move in a trajectory $z_1=b_1 + u_1 \tau$ and carry charge $Q_1$ so that it sources the background field
\[\label{eq:Particle1F}
\mathsf{\tilde A}_1^\mu(q)=-\frac{Q_1}{q^2}\,e^{iq\cdot b_1}\,\hat\delta(q\cdot u_1)\,u_1^\mu ,\qquad
f_{1,\mu\nu}=\frac{iQ_1}{q^2}e^{iq\cdot b_1}\hat\delta(q\cdot u_1)\,(q_\mu u_{1\nu}-q_\nu u_{1\mu}).
\]
where we have taken the gauge field to be in Lorenz gauge. Of course, our equations depend solely on the gauge invariant $f_{\mu\nu}$. Note that it is in principle straightforward to add spin to particle 1 via the substitution $u_1^\mu\to{\exp(-i\mathsf a_1*q)^\mu}_\nu u_1^\nu$, but we elect not to do so for convenience. Substituting eq.\eqref{eq:Particle1F} into equations \eqref{eq:dM}, \eqref{eq:du}, \eqref{eq:da}, and \eqref{eq:hidden} yields our desired worldline deflections.  
 Finally, we must remember that we solved our equations of motion for a single Fourier mode $q$. We therefore need to integrate the entire expression over $q$ to obtain our waveform result. Doing this and introducing the variables $q_1=-q$, $q_2=q-k$, we obtain the waveshape 
 \[\label{classwshap}
 \alpha_\eta(k)=i\!\int\!\hat \dd^4q_1\hat \dd^4q_2\,\hat\delta^4(k+q_1+q_2)\,\hat\delta(u_1\cdot q_1)\hat\delta(u_2\cdot q_2)\,
e^{-ib_1\cdot q_1}e^{-ib_2\cdot q_2}\;\frac{1}{q_1^2}\;\mathcal W\Big|_{q=-q_1,\ \mathsf{\tilde A}\to \mathsf{\tilde A}_1}.
\]
where the subscript $\mathsf{\tilde A} \rightarrow \mathsf{\tilde A}_1$ denotes the substitution of the background field \eqref{eq:Particle1F} sourced by particle 1, with its factor $e^{iq\cdot b_1}\hat\delta(q\cdot u_1)/q^2$ absorbed into the measure. With this normalisation $\mathcal W$ carries all the charge dependence: $\mathcal W_{\rm tree}=\mathcal O(Q_1Q_2^2)$, the same order as the tree-level five-point amplitude.  
The explicit expression for $\mathcal{W}$ is most compactly expressed by introducing the variable
\[
n^\mu\equiv u_1^\mu+\frac{\gamma}{\omega}\,q_1^\mu ,\qquad n\cdot u_2=0 ,
\label{eq:n}
\]
where $\gamma=u_1 \cdot u_2$ and working in the gauge where $\varepsilon^*\cdot u_2=0$. The vector $n^\mu$ lies in the direction of the Lorentz force exerted by particle 1 on particle 2, ie it is proportional to $\delta u_{2}^\mu$ at leading order. In terms of this vector we have
\[
\mathcal W^{(0)}_{\rm tree}&=-\frac{Q_1Q_2^2}{m_2}\,\varepsilon^*\!\cdot n ,\label{eq:Wt0}\\[4pt]
\mathcal W^{(1)}_{\rm tree}&=\frac{iQ_1Q_2^2}{m_2\,\omega}\Big[\omega\,\epsilon(\mathsf a_2,\varepsilon^*,k,n)-(k\cdot n)\,\epsilon(\mathsf a_2,\varepsilon^*,k,u_2)
-(\varepsilon^*\!\cdot q_1)\,\epsilon(\mathsf a_2,q_1,u_1,u_2)\\
&\hspace{6em}-(\mathsf a_2\cdot k)\,\epsilon(\varepsilon^*,q_1,u_1,u_2)+(\mathsf a_2\cdot\varepsilon^*)\,\epsilon(k,q_1,u_1,u_2)\Big],\\
\mathcal W^{(2)}_{\rm tree}&=\frac{Q_1Q_2^2}{m_2}\Big[-\tfrac12(\mathsf a_2\cdot q_2)^2\,\varepsilon^*\!\cdot n+\frac{k\cdot q_1}{\omega^2}\,B\Big],\\
B&=(\mathsf a_2\cdot k)\big[(\mathsf a_2\cdot n)(\varepsilon^*\!\cdot q_1)-(\mathsf a_2\cdot q_1)(\varepsilon^*\!\cdot n)\big]
+(\mathsf a_2\cdot\varepsilon^*)\big[(k\cdot n)(\mathsf a_2\cdot q_1)-(k\cdot q_1)(\mathsf a_2\cdot n)\big]\\
&\quad+\mathsf a_2^2\big[(\omega^2+k\cdot q_1)(\varepsilon^*\!\cdot n)-(k\cdot n)(\varepsilon^*\!\cdot q_1)\big],
\]
From here, it is a small step to evaluate the radiation reaction contribution to the waveshape. Indeed, it follows from substituting the radiation reaction deflections in eq. \eqref{eq:WaveformDeflection} so that 
\[\mathcal W\to\mathcal W_{\text{RR}}\equiv \mathcal W[\delta u_{2,{\text{RR}}},\delta z_{2,{\text{RR}}},\delta \ph_{2,{\text{RR}}},\delta\mathsf a_{2,{\text{RR}}},\delta \mathcal{M}_{2,{\text{RR}}}].\]
To do this, we need to first determine the radiation reaction equations of motion. We turn to this in the next section.

\section{Radiation reaction \`a la Sidney Coleman\label{sec: crr}}

In this section, we will calculate the classical radiation reaction following a method introduced by Coleman \cite{RM-2820-PR}. For this reason, we may refer to this method as CRR: Coleman/Classical Radiation Reaction. As the reader will see, the framework of this section belongs to classical field theory, yet we make use of QFT techniques such as tensor decomposition of integrals and dimensional regularisation to regularise the divergences that self-radiation brings with it.

\subsection{Regularised spinning multipoles for dissipation}
 
We start by considering  the $\sqrt{\rm Kerr}$ current, localised on a source worldline $z(\tau)$ and
written in momentum space as 
\[
\tilde J^\mu(k)=Q\!\int\!\mathrm d\tau\,e^{ik\cdot z(\tau)}\,\mathcal J^\mu(k,\tau),\,\,\,\,\,
\mathcal J^\mu(k,\tau)=u^\mu+i\frac{m}{\mathcal{M}}\,\epsilon^\mu(k,\ph,\sa)
 +\tfrac12(\sa\cdot k)\big((\sa\cdot k)\,\ph^\mu-(\ph\cdot k)\,\sa^\mu\big),
\]
The radiation reaction equations of motion require that we calculate the field strength at the location of the source. Specifically, we take the observation point $x$ to approach a specific point $z(0)$ on the worldline. We choose coordinates such that $z(0)$ is situated at the origin $x=0$, so that the self-field at this point is $F_{\mu \nu}(0)$, explicitly 
\[ \label{rrf}
&F_{\mu\nu}(0)=iQ\!\int\!\hd^Dk \!\int_{-\infty}^0\!\mathrm d\tau\,e^{ik\cdot z(\tau)}\,
\frac{k_{[\mu}\mathcal J_{\nu]}(k,\tau)}{k^2+i\epsilon k^0}.
\]
The self-field divergence is reflected in the large $k$ behaviour of the integrand, which grows quadratically in $k$. The $\sqrt{\rm Kerr}$ equations of motion \eqref{eq:TreeEOM} further require derivatives of $F_{\mu \nu}$ of the form:
\[\label{rrdf}
&\partial_{\alpha_1}\!\cdots\partial_{\alpha_n}F_{\mu\nu}(0)=(-i)^n\,iQ\!\int\!\hd^Dk\!\int_{-\infty}^0\!\mathrm d\tau\,
e^{ik\cdot z(\tau)}\,\frac{k_{\alpha_1}\cdots k_{\alpha_n}\,k_{[\mu}\mathcal J_{\nu]}(k,\tau)}{k^2+i\epsilon k^0},
\]
which are also UV divergent. As before, we extract the regular field strength by expanding the integrand for small $\tau$. Doing this leads to a proliferation of integrals, most of which are scaleless, and hence discarded in our dimensional regularisation setup.\footnote{Reference \cite{RM-2820-PR} reabsorbs the divergences into a noncovariant mass renormalisation scheme. Instead, for methods closer to us we point to \cite{DeAngelis:2026ymn, Galley:2010es,Kalin:2022hph}.} 

Consider an observer sitting near the particle with proper time $\tau=0$. The retarded field depends on worldline quantities at time $\tau_{\rm ret}$, which for an observer close to the worldline, is a small quantity (see figure \ref{fig:retarded-limit}). The worldline quantities at $\tau=\tau_{\rm ret}$ can therefore be expanded around $\tau=0$:
\begin{align}
    z^\mu(\tau_{\rm ret}) &= z^\mu(0)+ u^\mu\tau_{\rm ret}+\frac{1}{2}\dot{u}^\mu\tau_{\rm ret}^2+\frac{1}{3!}\ddot{u}^\mu\tau_{\rm ret}^3+\cdots \label{eq:rExpansion} \\
    u^\mu(\tau_{\rm ret}) &= u^\mu + \dot{u}^\mu\tau_{\rm ret}+ \frac{1}{2}\ddot{u}^\mu\tau_{\rm ret}^2+\cdots \label{eq:uExpansion} \\
    \sa^\mu(\tau_{\rm ret}) &= \sa^\mu + \dot{\sa}^\mu\tau_{\rm ret}+\frac{1}{2}\ddot{\sa}^\mu\tau_{\rm ret}^2+\cdots\ . \label{eq:aExpansion}
\end{align}
We will, for convenience, drop the subscript in $\tau_{\rm ret}$ in what follows. To expand about the straight line trajectory, we must power count such that $(k\cdot u)\tau\sim 1$, meaning that $\tau\sim k^{-1}$.

Once the expressions are expanded, the $\tau$ integral needs to be done. This is  easily done,
\begin{equation}
    \int_{-\infty}^0\!\mathrm{d}\tau\ e^{i(k\cdot u)\tau}\tau^m = - \frac{i^{m+1}m!}{(k\cdot u-i\epsilon)^{m+1}}\ .
\end{equation}
The $i\epsilon$ is added to ensure convergence as $\tau\to-\infty$. This only leaves the $k$ integral to be done, which in general are of the form 
\begin{equation}\label{eq:GeneralkIntegral}
    I^{\mu_1\cdots\mu_\ell}_{m} =\int\!\hd^Dk\ \frac{k^{\mu_1}\cdots k^{\mu_\ell}}{(k^2+i\epsilon k^0)(k\cdot u-i\epsilon)^{m+1}} \ . 
\end{equation}
By Lorentz invariance, these integrals must be functions solely of $u^2=1$, and thus any dimensionful integral must vanish. This limits
\begin{equation}\label{eq:DimensionlessCondition}
    D + \ell - 2 - (m+1) = 0\ ,
\end{equation}
to obtain non-zero integrals, cutting down greatly on the number of terms. In particular, it ensures that we can truncate the expansions in \eqref{eq:rExpansion}--\eqref{eq:aExpansion} to some finite order in $\tau$, since the number of $k^\mu$ in the numerator is generally fixed (when working at a fixed order in spin) and can only grow from expanding $\exp(ik\cdot z(\tau))$. But
\begin{equation}
    k\cdot z(\tau)-(k\cdot u)\tau = \frac{1}{2}k\cdot \dot{u} \tau^2+\cdots \propto k^{-1}(1+\cdots)\ ,
\end{equation}
so expanding the exponential always decreases the mass dimension of the integral. Therefore, eventually  all integrals beyond some order in $\tau$ will always be dimensionful. In fact, since eventually we want to work in $D=4$, from \eqref{eq:DimensionlessCondition} we can see that keeping all terms up to $\tau^{n+s+2}$ is enough for \eqref{rrf} expanded to the $s$th order in spin.

The remaining dimensionless integrals can be expressed, via PV reduction, in terms of the scalar dimensionless integral
\begin{equation}\label{eq:JDIntegral}
    J_D = \int\! \hd^D k\ \frac{1}{(k^2+i\epsilon k^0)(k\cdot u-i\epsilon)^{D-2}}\ .
\end{equation}
Evaluated for $D=4$, the integral is  nonzero if we perform something similar to a classical \emph{cut} over the photon propagator, as pointed out in \cite{Elkhidir:2023dco}. One finds
\begin{equation}
    J_4 = \int\!\hat\dd^4 k\ \frac{1}{(k^2+i\epsilon k^0)(k\cdot u-i\epsilon)^2} \xrightarrow[\text{cut}]{} \frac{1}{2}\int\!\hat\dd^4 k\ {\delta}(k^2)\delta'(u\cdot k)= \frac{1}{4\pi}\ ,
\end{equation}
which is what we will explicitly use for the waveshape evaluation.
The general form of the surviving tensor integral in \eqref{eq:GeneralkIntegral} is instead
\begin{equation}\label{formula}
    I^{\mu_1\dots \mu_\ell}_{D-3+\ell} = \int\! \hat\dd^Dk\ \frac{k^{\mu_1}\cdots k^{\mu_\ell}}{(k^2+i\epsilon k^0)(k\cdot u-i\epsilon)^{D-2+\ell}} = 2^\ell\, \frac{\Gamma\left(\ell+\frac{D-2}{2}\right)\Gamma(D-2) }{\Gamma\left(\frac{D-2}{2}\right)\Gamma(\ell+D-2) } J_{D} u^{\langle\mu_1}\cdots u^{\mu_\ell\rangle},
\end{equation}
where $u^{\langle\mu_1}\cdots u^{\mu_\ell\rangle}$ denotes the symmetric trace-free (STF) product of $u^{\mu_1}\cdots u^{\mu_\ell}$. For completeness, a few lower-rank examples are
\begin{align}
    I^{\mu}_{D-2} &= J_Du^\mu, \\
    I^{\mu\nu}_{D-1} &= J_D\left(\frac{D}{D-1}u^\mu u^\nu-\frac{1}{D-1}\eta^{\mu\nu}\right), \\ 
 I^{\mu\nu\rho}_{D} & = J_D\left(\frac{D+2}{D-1}u^\mu u^\nu u^\rho-\frac{1}{2(D-1)}\eta^{(\mu\nu}u^{\rho)}\right),\,
\end{align}
and  the special case of $D=4$ one has 
\begin{equation}
    \left.I^{\mu_1\dots \mu_\ell}_{\ell+1}\right|_{D=4} = \int\! \hat\dd^4k\ \frac{k^{\mu_1}\cdots k^{\mu_\ell}}{(k^2+i\epsilon k^0)(k\cdot u-i\epsilon)^{\ell+2}} =\frac{2^\ell}{4\pi(\ell+1)} u^{\langle\mu_1}\cdots u^{\mu_\ell\rangle}\, .
\end{equation}

It is more convenient to expand in spin and work out each order separately; we will denote the $n^\text{th}$ spin order with a superscript $(n)$. Once all the quantities are expanded, we obtain expressions for the renormalised field strengths in terms of $u$ and $\sa$ and their time derivatives.
Notice that each factor with time derivatives (eg. $\dot u, \ddot\sa\dots$) has to be evaluated on the corrected trajectories, starting  at tree-level with \eqref{eq:du}, \eqref{eq:da}, and \eqref{eq:hidden}. As a consequence, each such factor scales as $\mathcal O(Q_1Q_2)$. On the other hand, undifferentiated quantities can be evaluated on their background values. This allows us to separate out contributions by how distributed their time derivatives are: the leading contributions, with each term concentrating all the time derivatives on a single symbol, are known as Schott terms \cite{schott1912electromagnetic}. These terms can often be disregarded in certain computations, such as the impulse, since they are total derivative terms and vanish when integrated over the entire trajectory. The non-Schott terms first enter in the calculation of the impulse at the two-loop order $\mathcal{O}(Q_1^2 Q_2 ^4)$, see for instance \cite{Kosower:2018adc, Saketh:2021sri, Bern:2021xze}.   Schott terms however, even though they are total derivatives,  \emph{shall not} be thrown away whenever  the $\dd\tau$ integral is over an incomplete range, such as in \eqref{eq:dinttau} for the computation of $\delta u_2(\tau)$: it is precisely these local terms that are needed for the calculation of the waveform \cite{Elkhidir:2023dco}. With this in mind, the renormalised quantities that will be relevant to $\mathcal O(\sa^2)$ radiation reaction are found to be
\begingroup
\allowdisplaybreaks
\begin{align}\label{ff1}
    F^{(0)}_{\mu\nu} &= \frac{2}{3}QJ\ddot{u}_{[\mu}u_{\nu]}\ ,\\
    F^{(1)}_{\mu\nu} &=\frac{2}{3}QJ\epsilon_{\mu\nu}(u,\dddot{\sa})-\frac{2}{3}QJ\epsilon_{\mu\nu}(\sa,\dddot{u})+QJ u_{[\mu}\epsilon_{\nu]}(\sa,u,\dddot{u}) + \mathcal O(Q^2)\ , \\
    F^{(2)}_{\mu\nu} &= \frac{2}{5}QJ \sa\cdot \ddddot{u} u_{[\mu}\sa_{\nu]} + \frac{2}{15}QJ \sa^2 \ddddot{u}_{[\mu}u_{\nu]} + \mathcal O(Q^2)\ , \\
    \partial_{\alpha}F^{(0)}_{\mu\nu}&= \frac{1}{3} QJ\eta_{\alpha[\mu}\dddot{u}_{\nu]} + QJu_\alpha \dddot{u}_{[\mu}u_{\nu]} + \mathcal O(Q^2)\ ,\\
    \partial_\alpha F^{(1)}_{\mu\nu}&= \frac{1}{15}QJ
    \ddddot{u}_{[\mu}\epsilon_{\nu]\alpha}(\sa,u) +\frac{1}{3}QJ u_{[\mu}\epsilon_{\nu]\alpha}(\sa,\ddddot{u})  -\frac{1}{3}QJ     u_{[\mu}\epsilon_{\nu]\alpha}(u,\ddddot{\sa}) \nn\\*
    &\quad -\frac{4}{15}QJ \eta_{\alpha[\mu}\epsilon_{\nu]}(\sa,u,\ddddot{u}) +\frac{8}{5}QJ u^\alpha u_{[\mu}\epsilon_{\nu]}(\sa,u,\ddddot{u}) -\frac{2}{15}QJ \ddddot{u}_{\alpha}\epsilon_{\mu\nu}(\sa,u) \nn \\*
    &\quad  -\frac{2}{3}QJ u_\alpha\epsilon_{\mu\nu}(\sa,\ddddot{u}) +\frac{2}{3}QJ u_\alpha\epsilon_{\mu\nu}(u,\ddddot{\sa})  + \mathcal O(Q^2)\ , \\
    \partial_{\alpha}\partial_{\beta} F_{\mu\nu}^{(0)} &= \frac{1}{15}QJ \ddddot{u}_{(\alpha}\eta_{\beta)[\mu}u_{\nu]}+\frac{1}{3}QJ u_{(\alpha}\eta_{\beta)[\mu}\ddddot{u}_{\nu]}-\frac{4}{15}QJ\eta_{\alpha\beta}\ddddot{u}_{[\mu}u_{\nu]}\nn\\*
    &\quad-\frac{8}{5}QJu_\alpha u_\beta  u_{[\mu}\ddddot{u}_{\nu]} + \mathcal O(Q^2)\ ,\label{ff3}
\end{align}
\endgroup
where   $D=4$ simply corresponds to inserting  $J=J_4=1/4\pi$ in the relations above.

A few comments are now in order. We begin with two observations. First, if one tried to recompute the regularised field starting from  \eqref{rrf} but in the case of a charged and massless scalar one would find $F^{\mu\nu}\equiv0$ identically to all orders. This is a conclusion that was reached independently  in \cite{Lechner:2014kua}.\footnote{We thank Michael Saavedra for pointing this out to us.} From our point of view this is a consequence of the tensor decomposition in \eqref{eq:GeneralkIntegral} with  $u^2=0=\sa^\mu$: then  there is simply no nonzero Lorentz-invariant structure that the integrals can evaluate to. Secondly,  we find that if one tries to regularise a scalar messenger mediator with field
\begin{equation}\label{rrs}
    \phi (x) = Q\int\! \hd^Dk\ e^{-ik\cdot x}\int_{-\infty}^0\! \mathrm{d}\tau\ e^{ik\cdot z(\tau)}\frac{1}{k^2+i\epsilon k^0}\ ,
\end{equation}
one would find radiation reaction equations that are identical in form to the nonspinning ALD ones of electrodynamics, equation \eqref{eq:ald-equation}. This fact can be explained by dimensional analysis, Lorentz invariance and the fact that we need to have $u\cdot a=0$ to preserve on-shellness in the sense of $u^2=1$.  Next, we would like to stress  that  our method is indeed  valid beyond the Schott-terms sector approximation, given above in \eqref{ff1}--\eqref{ff3}. There, we simply truncated the equations for brevity and in light of the upcoming waveshape computation, where the Schott terms \eqref{ff1}--\eqref{ff3} are all that is needed. In fact, subleading non-Schott contributions can be easily  computed with our framework: we leave these for appendix \ref{appendix subl} where we show how these both match and extend known results at linear-spin order of \cite{Jakobsen:2023tvm}. Finally a comment on higher-spin terms. In fact, we find that the inclusion of arbitrarily-high spinning multipoles is in principle immediate from the structure of the calculation. These are simply entailed by the higher rank-integrals in equation \eqref{eq:JDIntegral}, which are known. However, one also needs the current \eqref{current} and the EOMs, which are in turn fixed by the dynamical mass-function of equation  \eqref{eq:DynamicalMass}. Unfortunately,   $\mathcal{M}$ is  not yet fully determined. Hence, the unknown  subleading terms in $\mathcal{M}$ are eventually the only obstacle. Interestingly, constraining such contact terms is a highly nontrivial task and an active area of research itself, both in EM and gravity \cite{Bautista:2021wfy,Bautista:2022wjf,Scheopner:2023rzp, deFirmian:2026mln, Bautista:2024agp}.

\subsection{Spinning radiation reaction EOMs and on-shell EM waveshape}

Next, we can finally  proceed by substituting the field strength tensors \eqref{ff1}--\eqref{ff3} into the equations of motion \eqref{eq:TreeEOM}. This yields the sought-after radiation reaction equations 
\[\label{eq:CRR}
\dot p_2^\mu\big|_{\text{RR}}&=\frac{Q_2^2}{6\pi}\ddot u_2^\mu-\frac{Q_2^2}{120\pi}\,T_2^{\mu\nu}\,\ddddot u_{2\nu}+\frac{Q_2^2}{6\pi}\,u_2^\mu\,(\mathsf a_2\cdot\ddddot{\mathsf a}_2),
\qquad T_2^{\mu\nu}=3\mathsf a_2^\mu\mathsf a_2^\nu-\mathsf a_2^2\eta^{\mu\nu},\\
(u_2-\ph_2)^\mu\big|_{\text{RR}}&=\frac{Q_2^2}{24\pi m_2}\big(3\mathsf a_2^2\eta^{\mu\nu}-5\mathsf a_2^\mu\mathsf a_2^\nu\big)\dddot u_{2\nu},\\
\dot{\mathsf a}_2^\mu\big|_{\text{RR}}&=-\frac{Q_2^2}{6\pi m_2}\Big[u_2^\mu\,(\mathsf a_2\cdot\ddot u_2)+\epsilon^\mu(u_2,\mathsf a_2,\dddot{\mathsf a}_2)\Big], 
\]
valid to order $\mathcal{O}(\sa^2)$.
The equation above gives the Schott terms in \eqref{mainn}, adding the non-total-derivative terms in the appendix \ref{appendix subl}, equation \eqref{eq:NLOEOMp}, one obtains  \eqref{mainn} in full. \eqref{eq:CRR} is, importantly, also truncated at linear order in the external field strength.

At this point, let us comment on potential issues arising from  higher-time derivatives on the RHS of the radiation reaction EOMs \eqref{eq:CRR}. In fact, these are known to lead to exponentially growing (runaway) solutions or causality-violating pre-acceleration. A perturbative solution to this is provided by \emph{reduction of order}, leading to the well-known  Landau-Lifshitz (LL) equation \cite{Landau:1975pou}. This equation  does not suffer from these pathologies, but ceases to be exact in the coupling. 
The LL form can simply be achieved  by replacing higher-order time derivatives on the RHS of \eqref{eq:CRR} with external fields $F_{\text{ext}}$, via the tree-level equations of motion.  Thus, linearising everything  one obtains immediately the spinning Landau-Lifshitz equations, with $F^{\,\text{ext}}_{\mu\nu}\equiv F_{\mu\nu}$ 
\[\label{eq:LLspin}
\dot p^\mu_2\big|_{\rm RR}&=\frac{Q^3_2}{6\pi m_2}\,u_2\cdot \partial \Big[F^\mu{}_\nu u^\nu_2
+P^{\mu\nu}\partial_\nu\Big({}^\star\! F_{\rho\sigma}\mathsf a^\rho_2 u^\sigma_2-\tfrac12\,\sa_2\cdot\partial\, F_{\rho\sigma}\mathsf a^\rho_2 u_2^\sigma\Big)\Big]\\
&\quad-\frac{Q^3_2}{120\pi m_2}\big(3\mathsf a^\mu_2\mathsf a^\nu_2-\mathsf a^2_2\eta^{\mu\nu}\big)\, (u_2\cdot \partial)^3F_{\nu\lambda}u^\lambda_2,\\
(u_2-\hat p_2)^\mu\big|_{\rm RR}&=\frac{Q^3_2}{24\pi m_2^2}\big(3\mathsf a^2_2\eta^{\mu\nu}-5\mathsf a^\mu_2\mathsf a^\nu_2\big)\,(u_2\cdot \partial)^2 F_{\nu\lambda}u^\lambda_2,\\
\dot{\mathsf a}_2^\mu\big|_{\rm RR}&=-\frac{Q^3_2}{6\pi m_2^2}\Big[u^\mu_2\,u_2\cdot \partial\,\Big(F_{\rho\sigma}\mathsf a^\rho _2 u^\sigma_2 +\sa_2\cdot \partial\,\,{}^\star\! F_{\rho\sigma}\mathsf a^\rho_2 u^\sigma_2\Big)
+\epsilon^\mu{}_{\sigma}(u_2,\sa_2) (u_2\cdot \partial)^2 F^\sigma{}_\lambda\mathsf a^\lambda_2\Big],\\
m_2\,\dot u^\mu_2\big|_{\rm RR}&=\frac{Q^3_2}{6\pi m_2}\,u_2\cdot \partial\,\Big[F^\mu{}_\nu u^\nu_2
+P^{\mu\nu}\partial_\nu\Big({}^\star\! F_{\rho\sigma}\mathsf a^\rho_2 u^\sigma_2-\tfrac12\,\sa_2\cdot \partial \,F_{\rho\sigma}\mathsf a^\rho_2 u^\sigma_2\Big)\Big]\\
&\quad-\frac{Q^3_2}{30\pi m_2}\big(7\mathsf a^\mu_2\mathsf a^\nu_2-4\mathsf a^2_2P^{\mu\nu}\big)\,(u_2\cdot\partial)^3F_{\nu\lambda}u^\lambda_2,
\]
where all derivatives act on the external field only, at this order $P^{\mu\nu}=\eta^{\mu\nu}-\hat p_2^\mu\hat p_2^\nu=\eta^{\mu\nu}-u_2^\mu u_2^\nu$, and the field strength above  can be taken as the one generated by particle 1, for instance: $F\equiv F_1 $. Note that the RHS now depends on $Q^3_2$ as the reduction is obtained via the leading EOMs for the variables $u^\mu_2(\tau),\sa^\mu_2(\tau)$. 
The equations above, \eqref{eq:CRR} with \eqref{eq:LLspin}, are one of the main results of our paper and, as a next step, we will use them to compute the dissipative waveshape in electrodynamics. Note that to do this, both  \eqref{eq:CRR} and \eqref{eq:LLspin} are  equivalent.

Computing the perturbative waveshape can be done by tracing the  steps   already detailed   in section \ref{sub:swaves}, where the tree-level spinning EM-waveshape was derived.\footnote{This is also a new result, to the best of our knowledge.} First, the tree-level data $\delta z_2, \delta \sa_2, \delta \ph_2, \delta u_2$ in equations \eqref{eq:du}--\eqref{eq:hidden} must be inserted into the RHS of \eqref{eq:CRR}.\footnote{This recursive process is indeed the same as the reduction of order that connects \eqref{eq:CRR} and \eqref{eq:LLspin}.}  Integrating \eqref{eq:CRR} with the tree-level data gives  $\mathcal{O} (Q_1Q_2^3)$ dynamics: $\delta z_{2,{\text{RR}}}, \delta \sa_{2,{\text{RR}}}, \delta \ph_{2,{\text{RR}}}, \delta u_{2,{\text{RR}}}$ etc. All we have to do is to insert the data inside 
\[\mathcal W\to\mathcal W[\delta u_{2,{\text{RR}}},\delta z_{2,{\text{RR}}},\delta \ph_{2,{\text{RR}}},\delta\mathsf a_{2,{\text{RR}}},\delta \mathcal{M}_{2,{\text{RR}}}]\equiv \mathcal W_{\text{RR}},\]
where $\mathcal{W}$ was  given in equation \eqref{eq:WaveformDeflection}. We finally find, truncating as usual  to quadratic spin order
\[\label{eq:CRRW}
\mathcal W^{(0)}_{\rm RR}&=-\frac{iQ_2^2\,\omega}{6\pi m_2}\,\mathcal W^{(0)}_{\rm tree}=\frac{iQ_1Q_2^4\,\omega}{6\pi m_2^2}\,\varepsilon^*\!\cdot n,\\
\mathcal W^{(1)}_{\rm RR}&=\frac{Q_1Q_2^4}{6\pi m_2^2}\Big[\omega\,\epsilon(\mathsf a_2,\varepsilon^*,k,n)-(k\cdot n)\,\epsilon(\mathsf a_2,\varepsilon^*,k,u_2)
-(\varepsilon^*\!\cdot q_1)\,\epsilon(\mathsf a_2,q_1,u_1,u_2)\Big],\\
\mathcal W^{(2)}_{\rm RR}&=-\frac{iQ_2^2\,\omega}{6\pi m_2}\,\mathcal W^{(2)}_{\rm tree}+\Delta,\\
\Delta&=\frac{iQ_1Q_2^4\,\omega}{\pi m_2^2}\Big[-\tfrac16(\mathsf a_2\cdot k)(\mathsf a_2\cdot q_1)\,\varepsilon^*\!\cdot n
+\Big(\frac{2\omega^2}{5}+\frac{k\cdot q_1}{6}\Big)(\mathsf a_2\cdot\varepsilon^*)(\mathsf a_2\cdot n)\\
&-\mathsf a_2^2\Big(\frac{3\omega^2}{10}\,\varepsilon^*\!\cdot n+\frac16\big[(k\cdot q_1)(\varepsilon^*\!\cdot n)-(k\cdot n)(\varepsilon^*\!\cdot q_1)\big]\Big)\Big],
\]
with $n^\mu$  from equation  \eqref{eq:n}. Note that the functions above are $\mathcal{O}(Q_1Q_2^4)$, the order of the one-loop sector, and give the waveshape once inserted in \eqref{classwshap}.

Although the results above are valid for $D=4$, the calculation can be repeated in other dimensions. For example, consider $D=6$: The scalar integral can be evaluated to $J_6=1/24\pi^2$. Since the change in dimension entails a change of mass dimension of the $\hat\dd^D k$ measure, a new set of dimensionless terms are picked up (those which were formerly, in the $D=4$ calculation, mass dimension $[M]^{-2}$). The resulting scalar ALD force is
\begin{equation}
    \dot p^\mu\bigr|_{{\text{RR}},D=6} = -\frac{Q^2}{30\pi^2}P^{\mu\nu}\left(\ddddot{u}_\nu+\frac{5}{2}\dot u^2 \ddot u_\nu+\frac{15}{2}\dot u\cdot \ddot u\dot u_\nu\right)\ .
\end{equation}
We have checked that this agrees with the results in \cite{Kosyakov:1999np}. This can be repeated for any $D=2n$. The issue with odd spacetime dimensions stems from the fact that the retarded Green's function has support \emph{inside} the lightcone, meaning that the motion depends on the history of the particle.  This breakdown is signalled by the fact that for odd $D$, $J_D$ can develop a logarithmic  divergence at large values of $|{\boldsymbol{k}}|$, whereas for even $D$ this does not happen. This is sometimes paraphrased by saying that Huygens' principle holds \emph{sharply} only in even spacetime dimensions\footnote{ Notoriously, this is not true in gravity:  when spacetime   curves there are tails even in $D=4$.} $D\geq 4$ \cite{PhysRevD.66.025016,Friedlander:2010eqa}.

The evaluation \eqref{eq:CRRW} concludes this section, which was devoted to a classical treatment of radiation reaction and spin.  In the next part of the paper we will be concerned with validating our results with an independent method, based on on-shell amplitudes and unitarity cuts.

\section{Spinning Compton cuts for dissipation of spinning bodies} \label{sec:ccut}

Here we shall be occupied with  confirming the validity of our spinning radiation reaction equations and waveshape. We will do this by comparing them to KMOC radiative observables, thus working with on-shell amplitude methods. Here, we will mostly follow and extend  the treatment of dissipation in terms of amplitude cuts, which was developed by some of the authors of this work in \cite{Elkhidir:2023dco} for the case of nonspinning dynamics.

\subsection{KMOC waveform essentials}

As we have seen in the previous section, the \emph{regular} self-field obtained determines a local correction to the particle's motion. During a scattering encounter, this correction also changes the radiation emitted by the particle: the ALD force perturbs its trajectory, and hence the current that sources the waveform. Let us consider the case where particle 2 is radiating and repeat once again the counting for clarity. Denoting the particles' charges by $Q_{1,2}$, then the leading acceleration of particle 2 under the influence of the field of particle 1 is of order $\mathcal O(Q_1Q_2)$, while the ALD correction to that acceleration will be $ \mathcal O (Q_1Q_2^3)$. The leading ALD contribution to the waveform is  then $\mathcal O(Q_1Q_2^4)$. This counting shows that radiation reaction consequently enters the waveform at the order described by one-loop five-point amplitudes.\footnote{On the other hand, it enters the momentum deflection only at two loops: $\mathcal O(Q_1^2 Q_2^4)$.} Led by this intuition, in this section we will exclusively be concerned with calculations in the sector $\mathcal O(Q_1Q_2^4)$, which is where radiation reaction first shows up in observables (the other gauge-invariant sector $\mathcal O(Q_2 Q_1^4)$ can be obtained by swapping labels).  
The next step is to then explain how this dissipative contribution can be identified directly from on-shell amplitudes and the KMOC formalism of ~\cite{Kosower:2018adc,Cristofoli:2021vyo}, without using the local self-force equation.

The KMOC formalism expresses classical scattering observables as
expectation values in suitably localised particle
states. For two incoming massive particles, we take
\begin{equation}
    |\psi\rangle
    =\sum_{\{c\}}\int \dd\Phi(p_1,p_2)\,
      \phi(p_1,p_2)e^{ib\cdot p_2}|p_1,p_2;\{c\}\rangle,
    \label{eq:kmoc-incoming-state}
\end{equation}
where $\phi_b(p_1,p_2)\equiv\phi(p_1,p_2)e^{ib\cdot p_2}$ specifies wavepackets centred on the incoming classical
momenta and separated by an impact parameter $b$. Above, $\{c\}$ represents a set of additional degrees of freedom of the colliding bodies that one may want to consider (for scalar electrodynamics $\{c\}=\emptyset$, but soon this index will be used to describe spin). 
In electromagnetism, the scattering of charged particles necessarily introduces radiation into the picture. In this setup, it can be shown that radiation is naturally encoded by the EM disturbance at some far-away point $x$ from the scattering event \cite{Cristofoli:2021vyo}
\begin{equation} 
  F_{\mu\nu}(x)
  =-\frac{1}{2\pi r}\,
  \Re\sum_\eta\int_0^\infty \hd\omega\,
  e^{-i\omega t_{\rm ret}}
  k_{[\mu}\varepsilon^{\eta}_{\nu]}(k)\,
  \alpha_\eta(k)
  +\mathcal O(r^{-2}),
  \qquad k^\mu=\omega(1,\hat{\bm x}),
  \label{eq:far-field-from-waveshape}
\end{equation}
with retarded time $t_{\rm ret}=x^0-r$, $r=|\vec x|$ and 
\[
 \alpha_\eta(k)
  \equiv
  \langle\psi|S^\dagger a_\eta(k)S|\psi\rangle= \int \dd\Phi(p'_1, p'_2, p_1, p_2) \, \phi_b^*(p_1', p_2') \phi_b(p_1, p_2) \braket{p_1' p_2' | S^\dagger a_\eta(k) S | p_1 p_2} ,
  \label{eq:waveshape-definition}
\]
which is what is referred to as ``waveshape". Above, $a_\eta(k)$ is the operator which annihilates an outgoing photon of momentum $k$ and helicity
$\eta$. The formula \eqref{eq:waveshape-definition} clearly reveals the deep relationship between observables and on-shell amplitudes: the radiative field at future null infinity is obtained from the on-shell Fourier transform of $\alpha_\eta(k)$, which is itself the expectation value of two  $S$-matrices. In fact, the waveshape was interpreted in  \cite{Caron-Huot:2023vxl}  as a \textit{generalised} amplitude, which is defined as an expectation value of one or more insertions of $S$. Importantly, as we are concerned with classical physics, the amplitudes entering the definition of  $\alpha_\eta(k)$ are computed in the classical-limit sense of \cite{Kosower:2018adc}. There, one expands in the regime where each massive momentum is much larger than the momenta of massless particles (messengers):
\[\label{scaling}
p^\mu=m u^\mu \gg q^\mu.
\]

Expanding the waveshape with $S=1+iT$, the authors of \cite{Elkhidir:2023dco} realised that at one loop $\alpha_\eta(k)$ acquires an imaginary part -- gauge invariant by itself --  which describes the effects of radiation reaction that enter the waveshape. This kernel is called ``CCut" as it is determined by a double-cutting rule that, at one loop,  isolates tree-level Compton amplitudes.  The fact that imaginary parts of amplitudes are related to cuts is nothing new: this is a known yet deep consequence of unitarity. What the authors of \cite{Elkhidir:2023dco} realised is that this can also be related to radiation reaction.  For our purposes, the CCut operation involves cutting the internal messenger propagator as well as the propagator of the massive line of the particle which is reacting to its own radiation. Pictorially this can be represented by the diagram 
\[
\begin{tikzpicture}[x=0.75pt,y=0.75pt,yscale=-1,xscale=1]

\tikzset{
pattern size/.store in=\mcSize, 
pattern size = 5pt,
pattern thickness/.store in=\mcThickness, 
pattern thickness = 0.3pt,
pattern radius/.store in=\mcRadius, 
pattern radius = 1pt}
\makeatletter
\pgfutil@ifundefined{pgf@pattern@name@_2hq71htmz}{
\pgfdeclarepatternformonly[\mcThickness,\mcSize]{_2hq71htmz}
{\pgfqpoint{0pt}{0pt}}
{\pgfpoint{\mcSize+\mcThickness}{\mcSize+\mcThickness}}
{\pgfpoint{\mcSize}{\mcSize}}
{
\pgfsetcolor{\tikz@pattern@color}
\pgfsetlinewidth{\mcThickness}
\pgfpathmoveto{\pgfqpoint{0pt}{0pt}}
\pgfpathlineto{\pgfpoint{\mcSize+\mcThickness}{\mcSize+\mcThickness}}
\pgfusepath{stroke}
}}
\makeatother

 
\tikzset{
pattern size/.store in=\mcSize, 
pattern size = 5pt,
pattern thickness/.store in=\mcThickness, 
pattern thickness = 0.3pt,
pattern radius/.store in=\mcRadius, 
pattern radius = 1pt}
\makeatletter
\pgfutil@ifundefined{pgf@pattern@name@_sm4n8myyw}{
\pgfdeclarepatternformonly[\mcThickness,\mcSize]{_sm4n8myyw}
{\pgfqpoint{0pt}{0pt}}
{\pgfpoint{\mcSize+\mcThickness}{\mcSize+\mcThickness}}
{\pgfpoint{\mcSize}{\mcSize}}
{
\pgfsetcolor{\tikz@pattern@color}
\pgfsetlinewidth{\mcThickness}
\pgfpathmoveto{\pgfqpoint{0pt}{0pt}}
\pgfpathlineto{\pgfpoint{\mcSize+\mcThickness}{\mcSize+\mcThickness}}
\pgfusepath{stroke}
}}
\makeatother
\tikzset{every picture/.style={line width=0.75pt}} 


\draw [color={rgb, 255:red, 208; green, 2; blue, 27 }  ,draw opacity=1 ] [dash pattern={on 4.5pt off 4.5pt}]  (473.3,13200) -- (473,13134) ;
\draw  [pattern=_2hq71htmz,pattern size=3.2249999999999996pt,pattern thickness=0.75pt,pattern radius=0pt, pattern color={rgb, 255:red, 0; green, 0; blue, 0}] (442.36,13168.57) .. controls (442.32,13175.98) and (436.22,13181.96) .. (428.74,13181.92) .. controls (421.25,13181.87) and (415.22,13175.83) .. (415.26,13168.42) .. controls (415.3,13161.01) and (421.4,13155.03) .. (428.88,13155.08) .. controls (436.37,13155.12) and (442.4,13161.16) .. (442.36,13168.57) -- cycle ;
\draw [color={rgb, 255:red, 0; green, 135; blue, 189 }  ,draw opacity=1 ][line width=1.5]    (438.97,13178.1) .. controls (446.52,13187.86) and (455.72,13190.11) .. (466.93,13190.16) ;
\draw [color={rgb, 255:red, 208; green, 2; blue, 27 }  ,draw opacity=1 ][line width=1.5]    (428.88,13155.08) .. controls (430.25,13126.66) and (452,13103) .. (477,13103) ;
\draw [color={rgb, 255:red, 208; green, 2; blue, 27 }  ,draw opacity=1 ][line width=1.5]    (396.46,13141.81) -- (418.14,13159.87) ;
\draw [color={rgb, 255:red, 0; green, 135; blue, 189 }  ,draw opacity=1 ][line width=1.5]    (417.19,13176.13) -- (394.79,13195.06) ;
\draw  [pattern=_sm4n8myyw,pattern size=3.2249999999999996pt,pattern thickness=0.75pt,pattern radius=0pt, pattern color={rgb, 255:red, 0; green, 0; blue, 0}] (503.35,13168.57) .. controls (503.39,13175.98) and (509.49,13181.96) .. (516.97,13181.92) .. controls (524.46,13181.87) and (530.49,13175.83) .. (530.45,13168.42) .. controls (530.41,13161.01) and (524.31,13155.03) .. (516.83,13155.08) .. controls (509.35,13155.12) and (503.31,13161.16) .. (503.35,13168.57) -- cycle ;
\draw [color={rgb, 255:red, 0; green, 135; blue, 189 }  ,draw opacity=1 ][line width=1.5]    (506.74,13178.1) .. controls (499.19,13187.86) and (490,13190.11) .. (478.79,13190.16) ;
\draw [color={rgb, 255:red, 0; green, 135; blue, 189 }  ,draw opacity=1 ][line width=1.5]    (528.53,13176.13) -- (550.93,13195.06) ;
\draw    (523.45,13157.25) .. controls (524.16,13155) and (525.63,13154.22) .. (527.88,13154.93) .. controls (530.13,13155.63) and (531.6,13154.86) .. (532.31,13152.61) .. controls (533.02,13150.36) and (534.49,13149.59) .. (536.74,13150.3) .. controls (538.99,13151) and (540.46,13150.23) .. (541.17,13147.98) .. controls (541.88,13145.73) and (543.36,13144.95) .. (545.61,13145.66) .. controls (547.86,13146.36) and (549.33,13145.59) .. (550.04,13143.34) -- (551.33,13142.67) -- (551.33,13142.67) ;
\draw    (466.33,13144) .. controls (464.53,13145.65) and (462.79,13145.62) .. (461.11,13143.91) .. controls (459.59,13142.18) and (457.96,13142.12) .. (456.23,13143.71) .. controls (454.62,13145.34) and (453,13145.42) .. (451.35,13143.95) .. controls (448.78,13143.56) and (447.46,13144.59) .. (447.41,13147.05) .. controls (447.66,13149.35) and (446.58,13150.66) .. (444.17,13150.99) .. controls (441.84,13151.2) and (440.77,13152.49) .. (440.94,13154.87) .. controls (441.07,13157.27) and (439.96,13158.55) .. (437.59,13158.72) -- (437.33,13159) ;
\draw    (508.33,13158.33) .. controls (505.99,13158) and (504.96,13156.62) .. (505.25,13154.2) .. controls (505.66,13151.91) and (504.7,13150.57) .. (502.37,13150.19) .. controls (500.1,13150) and (498.99,13148.8) .. (499.04,13146.59) .. controls (497.99,13144.3) and (496.43,13143.79) .. (494.35,13145.04) .. controls (492.42,13146.45) and (490.75,13146.22) .. (489.35,13144.37) .. controls (487.81,13142.5) and (486.1,13142.28) .. (484.23,13143.71) .. controls (482.33,13145.12) and (480.7,13144.89) .. (479.33,13143) -- (479.33,13143) ;

\draw (380,13130) node [anchor=north west][inner sep=0.75pt]    {$p_{1}$};
\draw (377,13194) node [anchor=north west][inner sep=0.75pt]    {$p_{2}$};
\draw (479,13094) node [anchor=north west][inner sep=0.75pt]    {$p'_{1}$};
\draw (555,13191) node [anchor=north west][inner sep=0.75pt]    {$p'_{2}$};
\draw (554,13134) node [anchor=north west][inner sep=0.75pt]    {$k$};
\draw (574,13164) node [anchor=north west][inner sep=0.75pt]    {$,$};
\draw (150,13154) node [anchor=north west][inner sep=0.75pt]    {$2\CCut_2 \mathcal{A}_{5,1}^{Q_1 Q_2^4}(p_1, p_2 \rightarrow p'_1, p'_2, k_\eta)=$};

\end{tikzpicture}
\]
where the label on $\CCut_i$ tells us about which massive line is cut. Note that we are using the fragment notation of \cite{Cristofoli:2021jas}: $\mathcal{A}_{\text{points, loops}}$.
In the end, at the order we are concerned with,  one can show that 
\[\label{eq:genericWaveshapeIm}
\waveshape|_\textrm{RR} = -\KMOCav{\int \hd^D q_1 \hd^D q_2 \, \hdelta(2 p_1 \cdot q_1) \hdelta(2 p_2 \cdot q_2) \,& \hdelta^D(k + q_1 + q_2) \,e^{-ib_1 \cdot q_1}e^{-ib_2 \cdot q_2} 
\\ & \times \CCut_2 \mathcal{A}_{5,1}^{Q_1 Q_2^4}(p_1, p_2 \rightarrow p'_1, p'_2, k_\eta)} \,,
\]
where the KMOC double brackets \cite{Kosower:2018adc} are defined as an average over the semiclassical wavepackets in \eqref{eq:kmoc-incoming-state}
\[
\KMOCav{\cdots}=\int \dd \Phi (p_1,p_2)|\phi(p_1,p_2)|^2(\,\cdots),
\]
note that the finer details of the wave packets $\phi$ are not needed explicitly: for us, they will simply enforce the scaling  in \eqref{scaling}. Keeping this in mind we will omit these brackets from now on. 

We conclude by noting that the imaginary part of the waveshape can also receive contributions from a  different type of iteration diagrams, see \cite{Elkhidir:2023dco, Caron-Huot:2023vxl}, but these are zero within the dissipative sector $\mathcal O(Q_1Q_2^4)$ we are considering.

\subsection{Cut-Compton amplitudes with spin for spinning radiation reaction}
So far we have seen how in  \cite{Elkhidir:2023dco} dissipative  \textit{scalar} radiation (radiation generated by a scalar particle's own field) was characterised in terms of the imaginary part of the  waveshape. What we will show next is that the generality of the CCut allows us to extend the formalism described above to include spinning effects in a very direct way; thus allowing us to characterize spinning radiation reaction from an amplitude standpoint.
We define the spinning Compton cut simply as  $\CCut_{i,\sa_i}$, where one now cuts over the $i^{\text{th}}$-spinning massive line:
\[\label{spinningccut}
\begin{tikzpicture}[x=0.75pt,y=0.75pt,yscale=-1,xscale=1]

 
\tikzset{
pattern size/.store in=\mcSize, 
pattern size = 5pt,
pattern thickness/.store in=\mcThickness, 
pattern thickness = 0.3pt,
pattern radius/.store in=\mcRadius, 
pattern radius = 1pt}
\makeatletter
\pgfutil@ifundefined{pgf@pattern@name@_2hq71htmz}{
\pgfdeclarepatternformonly[\mcThickness,\mcSize]{_2hq71htmz}
{\pgfqpoint{0pt}{0pt}}
{\pgfpoint{\mcSize+\mcThickness}{\mcSize+\mcThickness}}
{\pgfpoint{\mcSize}{\mcSize}}
{
\pgfsetcolor{\tikz@pattern@color}
\pgfsetlinewidth{\mcThickness}
\pgfpathmoveto{\pgfqpoint{0pt}{0pt}}
\pgfpathlineto{\pgfpoint{\mcSize+\mcThickness}{\mcSize+\mcThickness}}
\pgfusepath{stroke}
}}
\makeatother

 
\tikzset{
pattern size/.store in=\mcSize, 
pattern size = 5pt,
pattern thickness/.store in=\mcThickness, 
pattern thickness = 0.3pt,
pattern radius/.store in=\mcRadius, 
pattern radius = 1pt}
\makeatletter
\pgfutil@ifundefined{pgf@pattern@name@_sm4n8myyw}{
\pgfdeclarepatternformonly[\mcThickness,\mcSize]{_sm4n8myyw}
{\pgfqpoint{0pt}{0pt}}
{\pgfpoint{\mcSize+\mcThickness}{\mcSize+\mcThickness}}
{\pgfpoint{\mcSize}{\mcSize}}
{
\pgfsetcolor{\tikz@pattern@color}
\pgfsetlinewidth{\mcThickness}
\pgfpathmoveto{\pgfqpoint{0pt}{0pt}}
\pgfpathlineto{\pgfpoint{\mcSize+\mcThickness}{\mcSize+\mcThickness}}
\pgfusepath{stroke}
}}
\makeatother
\tikzset{every picture/.style={line width=0.75pt}} 


\draw [color={rgb, 255:red, 208; green, 2; blue, 27 }  ,draw opacity=1 ] [dash pattern={on 4.5pt off 4.5pt}]  (473.3,13200) -- (473,13134) ;
\draw  [pattern=_2hq71htmz,pattern size=3.2249999999999996pt,pattern thickness=0.75pt,pattern radius=0pt, pattern color={rgb, 255:red, 0; green, 0; blue, 0}] (442.36,13168.57) .. controls (442.32,13175.98) and (436.22,13181.96) .. (428.74,13181.92) .. controls (421.25,13181.87) and (415.22,13175.83) .. (415.26,13168.42) .. controls (415.3,13161.01) and (421.4,13155.03) .. (428.88,13155.08) .. controls (436.37,13155.12) and (442.4,13161.16) .. (442.36,13168.57) -- cycle ;
\draw [color={rgb, 255:red, 0; green, 135; blue, 189 }  ,draw opacity=1 ][line width=1.5]    (438.97,13178.1) .. controls (446.52,13187.86) and (455.72,13190.11) .. (466.93,13190.16) ;
\draw [color={rgb, 255:red, 208; green, 2; blue, 27 }  ,draw opacity=1 ][line width=1.5]    (428.88,13155.08) .. controls (430.25,13126.66) and (452,13103) .. (477,13103) ;
\draw [color={rgb, 255:red, 208; green, 2; blue, 27 }  ,draw opacity=1 ][line width=1.5]    (396.46,13141.81) -- (418.14,13159.87) ;
\draw [color={rgb, 255:red, 0; green, 135; blue, 189 }  ,draw opacity=1 ][line width=1.5]    (417.19,13176.13) -- (394.79,13195.06) ;
\draw  [pattern=_sm4n8myyw,pattern size=3.2249999999999996pt,pattern thickness=0.75pt,pattern radius=0pt, pattern color={rgb, 255:red, 0; green, 0; blue, 0}] (503.35,13168.57) .. controls (503.39,13175.98) and (509.49,13181.96) .. (516.97,13181.92) .. controls (524.46,13181.87) and (530.49,13175.83) .. (530.45,13168.42) .. controls (530.41,13161.01) and (524.31,13155.03) .. (516.83,13155.08) .. controls (509.35,13155.12) and (503.31,13161.16) .. (503.35,13168.57) -- cycle ;
\draw [color={rgb, 255:red, 0; green, 135; blue, 189 }  ,draw opacity=1 ][line width=1.5]    (506.74,13178.1) .. controls (499.19,13187.86) and (490,13190.11) .. (478.79,13190.16) ;
\draw [color={rgb, 255:red, 0; green, 135; blue, 189 }  ,draw opacity=1 ][line width=1.5]    (528.53,13176.13) -- (550.93,13195.06) ;
\draw    (523.45,13157.25) .. controls (524.16,13155) and (525.63,13154.22) .. (527.88,13154.93) .. controls (530.13,13155.63) and (531.6,13154.86) .. (532.31,13152.61) .. controls (533.02,13150.36) and (534.49,13149.59) .. (536.74,13150.3) .. controls (538.99,13151) and (540.46,13150.23) .. (541.17,13147.98) .. controls (541.88,13145.73) and (543.36,13144.95) .. (545.61,13145.66) .. controls (547.86,13146.36) and (549.33,13145.59) .. (550.04,13143.34) -- (551.33,13142.67) -- (551.33,13142.67) ;
\draw    (466.33,13144) .. controls (464.53,13145.65) and (462.79,13145.62) .. (461.11,13143.91) .. controls (459.59,13142.18) and (457.96,13142.12) .. (456.23,13143.71) .. controls (454.62,13145.34) and (453,13145.42) .. (451.35,13143.95) .. controls (448.78,13143.56) and (447.46,13144.59) .. (447.41,13147.05) .. controls (447.66,13149.35) and (446.58,13150.66) .. (444.17,13150.99) .. controls (441.84,13151.2) and (440.77,13152.49) .. (440.94,13154.87) .. controls (441.07,13157.27) and (439.96,13158.55) .. (437.59,13158.72) -- (437.33,13159) ;
\draw    (508.33,13158.33) .. controls (505.99,13158) and (504.96,13156.62) .. (505.25,13154.2) .. controls (505.66,13151.91) and (504.7,13150.57) .. (502.37,13150.19) .. controls (500.1,13150) and (498.99,13148.8) .. (499.04,13146.59) .. controls (497.99,13144.3) and (496.43,13143.79) .. (494.35,13145.04) .. controls (492.42,13146.45) and (490.75,13146.22) .. (489.35,13144.37) .. controls (487.81,13142.5) and (486.1,13142.28) .. (484.23,13143.71) .. controls (482.33,13145.12) and (480.7,13144.89) .. (479.33,13143) -- (479.33,13143) ;

\draw (380,13130) node [anchor=north west][inner sep=0.75pt]    {$p_{1}$};
\draw (377,13199) node [anchor=north west][inner sep=0.75pt]    {$p_{2},\sa_2$};
\draw (479,13094) node [anchor=north west][inner sep=0.75pt]    {$p'_{1}$};
\draw (555,13191) node [anchor=north west][inner sep=0.75pt]    {$p'_{2},\sa_2'$};
\draw (554,13134) node [anchor=north west][inner sep=0.75pt]    {$k$};
\draw (574,13164) node [anchor=north west][inner sep=0.75pt]    {$,$};
\draw (130,13154) node [anchor=north west][inner sep=0.75pt]    {$2 \CCut_{2,\sa_2} \mathcal{A}_{5,1}^{\sa_2 Q_1 Q_2^4}(p_1, p_2 \rightarrow p'_1, p'_2, k_\eta)=$};

\end{tikzpicture}
\]
where $\sa_2^\mu=\mathsf{S}^\mu_2/m_2$ is a spacelike vector describing the classical spin (in units of length) of particle 2, just like in section \ref{sec: crr}. 
Next, a few comments on classical spin are in order. 

There are different formulations of classical spin. For instance, within the KMOC framework \cite{Maybee:2019jus, Aoude:2021oqj}  it is natural  to describe classical spin    in terms of the expectation value of the Pauli-Lubanski pseudovector operator $\mathbb{\hat W}^\mu$: $\langle \psi| \mathbb{\hat W}^\mu_i| \psi\rangle$,
where now $|\psi\rangle$ is a KMOC state of the form \eqref{eq:kmoc-incoming-state} where now the summation   in $\sum_{\{c\}}|p_2, \{c\}\rangle$ (if we take only particle 2 to have spin) is   a sum over large, coherent, representations of $SU(2)$ and $\{c\}\equiv s$ is the spin quantum number.  The classical limit of spin arises when the ``size" of this representation grows to infinity with the  classical scaling \cite{Aoude:2021oqj, Arkani-Hamed:2019ymq}
\[
s\to \infty,\,\,\,\,\hbar\to 0,\,\,\,\, \hbar \,s \sim \mathcal O (1),
\]
so in this sense we can define the classically finite spin vector  as \cite{Maybee:2019jus} 
\[
 \sa^\mu_2 = \frac{1}{m_2^2}   \langle \psi| \mathbb{\hat W}^\mu_2| \psi\rangle.
\]
Spinning amplitudes arise as $iT$ expectation values over KMOC spinning states that enjoy the classical scaling described above. It is precisely these kinds of amplitudes that enter the definition of the Compton cut given above in \eqref{spinningccut}.

It is then clear at this point that the crucial building block of the CCut is the spinning Compton amplitude at tree level. This is an object which itself received a great deal of attention in the last few years. One main reason for this is that, writing the amplitude as a Taylor expansion around spinning multipoles as 
\[
\mathcal{A}^{\sa_2}_{4,0}=\mathcal J^{(0)}+\mathcal J^{(1)}\cdot  \sa_2 +\mathcal J^{(2)}\cdot  \sa_2\cdot \sa_2 +\mathcal J^{(3)}\cdot  \sa_2\cdot \sa_2\cdot \sa_2+\cdots
\]
one finds that for multipoles $\sa_2^{n>2}$, then   $\mathcal J^{(n>2)}$  presents spurious singularities (i.e. poles that do not correspond  to physical factorization channels)  that need extra care \cite{Arkani-Hamed:2017jhn, Guevara:2018wpp}. Finding a way to eliminate the unphysical poles by constraining the amplitudes  with QFT principles or black hole perturbation theory methods has motivated its own independent research program, see for instance \cite{Bautista:2021wfy, Bautista:2022wjf, Cangemi:2023bpe,Cangemi:2023ysz, Cangemi:2022bew, Chen:2024mlx,Bjerrum-Bohr:2025bqg,Bjerrum-Bohr:2026fhx,Bjerrum-Bohr:2023jau, Vazquez-Holm:2025ztz, Aoude:2022thd,Aoude:2023dui,Aoude:2023vdk, Aoude:2022trd, Bern:2022kto, Ochirov:2022nqz, Bautista:2023sdf, Kim:2026fds} where the same issue is also discussed in a gravitational setup. Note also that this drawback is essentially the same that appeared earlier in section \ref{sec: crr}, where we discussed ambiguities in higher-order spin operators of $\mathcal{M}$: it is now only reformulated in terms of the classical limit of spinning amplitudes.  Eventually, in this work we shall not be concerned with these issues too much  as we can simply choose to restrict to $\sa^{n<3}_2$. Here, spinning Compton amplitudes are under control yet rich enough to allow us to test our ideas nontrivially. Furthermore we anticipate here that higher spinning multipoles can be incorporated algorithmically in our formalism, if one were to decide to include them. This can certainly be explored in a future work. We will comment more on this point in the concluding section.

\subsection{Spinning CCut evaluation}
Let us then compute the CCut now.
In what follows we will  take only particle 2 to be spinning, the particle which is reacting,  so 
\[
\sa_1^\mu\equiv0.
\]
This is to keep the proliferation of long formulas under control and  can be done without losing generality: the source field (particle 1, say) is for us a mere spectator: its only role  is to furnish the initial acceleration that allows the two-body radiation to be eventually emitted.\ On the other hand, it is in principle less obvious how the self-interacting dynamics of particle 2 will affect its spin.  Moreover, including spin on the other particle can be actually done without effort at this order, we will comment more later.

The spinning tree-level, four-point, Compton amplitude we need can be found for instance in \cite{Bern:2023ity} (See the tree amplitudes sec III.B there, with the $\sqrt{\text{Kerr}}$ choice $C_i=1, \,\,\,D_i=0$) or in \cite{Scheopner:2023rzp} (Eq. 5.34, still with $C_i=1, \,\,\,D_i=0$).  
For completeness, we report it here as well. Writing 
\[\label{compspin}
\mathcal{A}^{\sa_2}_{4,0}(p_2, k_1\to p_2, k_2)&
=2Q_2^2\varepsilon^\mu _\eta (k_1)\varepsilon^{\nu *}_{\eta'} (k_2) \mathcal J_{\mu\nu}^{\sa_2}(p_2, k_1\to p_2, k_2),\]
we have, up to spin-squared order
\[\mathcal  J_{\mu\nu}^{\sa_2}(p_2, k_1\to p_2, k_2) \equiv   
\mathcal J^{(0)}_{\mu\nu}+\mathcal J^{(1)}_{\mu\nu,\lambda}{\sa_2}^\lambda+\mathcal J^{(2)}_{\mu\nu,\lambda\xi}\sa^\lambda_2 \sa^\xi_2,
\]
together with the scalar current  defined by 
\[
\mathcal J^{(0)}_{\mu\nu}=-
\eta_{\mu\nu}
+\frac{u_{2\mu}k_{1\nu}-k_{2\mu}u_{2\nu}}
{u_2\cdot k_1}
+
\frac{k_1\cdot k_2 \,u_{2\mu}u_{2\nu}}
{(u_2\cdot k_1)^2},
\]
 the spin-orbit one 
\[
\mathcal  J^{(1)}_{\mu\nu,\lambda}
=&
    \frac{iu_2^\alpha}{u_2\cdot k_1}
\Big(
k_{2\mu}\epsilon_{\nu\rho\alpha\lambda}
(k_1+k_2)^\rho
-k_{1\nu}\epsilon_{\mu\rho\alpha\lambda}
(k_1+k_2)^\rho 
+k_1\cdot k_2\,\epsilon_{\mu\nu\alpha\lambda}
-\eta_{\mu\nu}\epsilon_{\rho\sigma\alpha\lambda}
k_2^\rho k_1^\sigma
\Big)
\\
&\qquad\hspace{6cm}
-i\frac{k_1\cdot k_2}{(u_2\cdot k_1)^2}
\Big(
u_{2\nu}\epsilon_{\mu\rho\alpha\lambda}k_1^\rho
+u_{2\mu}\epsilon_{\nu\rho\alpha\lambda}k_2^\rho
\Big) ,
\]
and finally 
\[
\mathcal J_{\mu\nu,\lambda\xi}^{(2)}
=\,& 
(k_1\!+k_2)_\lambda(k_1\!+k_2)_\xi
\mathcal J^{(0)}_{\mu\nu}\!
+
\frac{k_1\cdot k_2}{(u_2\cdot k_1)^2}\!
\Bigg[
k_{1\lambda}k_{2\xi}\eta_{\mu\nu}\!
+k_1\cdot k_2\eta_{\mu\lambda}\eta_{\nu\xi}
\!-k_{2\xi}k_{1\nu}\eta_{\mu\lambda}\!\!
-k_{1\lambda}k_{2\mu}\eta_{\nu\xi}
\\
&\hspace{0.7cm}
+\eta_{\lambda\xi}
\Big(
(k_{1\nu}-u_2\cdot k_1\,u_{2\nu})
(k_{2\mu}+u_2\cdot k_1\,u_{2\mu})
-(\eta_{\mu\nu}-u_{2\mu}u_{2\nu})
(k_1\cdot k_2+(u_2\cdot k_1)^2)
\Big)
\Bigg],
\]
is the spin-squared part. In all the spin sectors above, we wrote $p_2^\mu=m_2u_2^\mu$ and simplified a bit using momentum conservation and that $\sa_2\cdot u_2=0$. These amplitudes are the ones of reference \cite{Bern:2023ity} with the necessary notational changes. Above, we have also expressed spin in terms of the spin vector $\sa_2^\mu$, which relates to the spin tensor in \cite{Bern:2023ity} through  $\sS^{\mu\nu}_2=m_2\epsilon^{\mu\nu\rho\sigma}u_{2\rho}\sa_{2\sigma}$. Note that as $\sa^\mu_2$ is a pseudovector, every $\mathcal{O}(\sa_2)$ term is always accompanied by a Levi-Civita pseudotensor by parity conservation.

At this point all we have to do is to insert the Compton amplitude \eqref{compspin} into the cut. In fact, the latter actually depends on two copies of $\mathcal{A}_{4,0}^{\sa_2}$: this can be seen by further resolving the $1/q_1^2$ propagator with a unitarity cut \cite{Elkhidir:2023dco} leading to the following expression:

\[
\begin{tikzpicture}[x=0.75pt,y=0.75pt,yscale=-1,xscale=1]
 
\tikzset{
pattern size/.store in=\mcSize, 
pattern size = 5pt,
pattern thickness/.store in=\mcThickness, 
pattern thickness = 0.3pt,
pattern radius/.store in=\mcRadius, 
pattern radius = 1pt}
\makeatletter
\pgfutil@ifundefined{pgf@pattern@name@_n879p6llu}{
\pgfdeclarepatternformonly[\mcThickness,\mcSize]{_n879p6llu}
{\pgfqpoint{0pt}{0pt}}
{\pgfpoint{\mcSize+\mcThickness}{\mcSize+\mcThickness}}
{\pgfpoint{\mcSize}{\mcSize}}
{
\pgfsetcolor{\tikz@pattern@color}
\pgfsetlinewidth{\mcThickness}
\pgfpathmoveto{\pgfqpoint{0pt}{0pt}}
\pgfpathlineto{\pgfpoint{\mcSize+\mcThickness}{\mcSize+\mcThickness}}
\pgfusepath{stroke}
}}
\makeatother

 
\tikzset{
pattern size/.store in=\mcSize, 
pattern size = 5pt,
pattern thickness/.store in=\mcThickness, 
pattern thickness = 0.3pt,
pattern radius/.store in=\mcRadius, 
pattern radius = 1pt}
\makeatletter
\pgfutil@ifundefined{pgf@pattern@name@_izjrheo3s}{
\pgfdeclarepatternformonly[\mcThickness,\mcSize]{_izjrheo3s}
{\pgfqpoint{0pt}{0pt}}
{\pgfpoint{\mcSize+\mcThickness}{\mcSize+\mcThickness}}
{\pgfpoint{\mcSize}{\mcSize}}
{
\pgfsetcolor{\tikz@pattern@color}
\pgfsetlinewidth{\mcThickness}
\pgfpathmoveto{\pgfqpoint{0pt}{0pt}}
\pgfpathlineto{\pgfpoint{\mcSize+\mcThickness}{\mcSize+\mcThickness}}
\pgfusepath{stroke}
}}
\makeatother

 
\tikzset{
pattern size/.store in=\mcSize, 
pattern size = 5pt,
pattern thickness/.store in=\mcThickness, 
pattern thickness = 0.3pt,
pattern radius/.store in=\mcRadius, 
pattern radius = 1pt}
\makeatletter
\pgfutil@ifundefined{pgf@pattern@name@_vkibh8r0o}{
\pgfdeclarepatternformonly[\mcThickness,\mcSize]{_vkibh8r0o}
{\pgfqpoint{0pt}{0pt}}
{\pgfpoint{\mcSize+\mcThickness}{\mcSize+\mcThickness}}
{\pgfpoint{\mcSize}{\mcSize}}
{
\pgfsetcolor{\tikz@pattern@color}
\pgfsetlinewidth{\mcThickness}
\pgfpathmoveto{\pgfqpoint{0pt}{0pt}}
\pgfpathlineto{\pgfpoint{\mcSize+\mcThickness}{\mcSize+\mcThickness}}
\pgfusepath{stroke}
}}
\makeatother

 
\tikzset{
pattern size/.store in=\mcSize, 
pattern size = 5pt,
pattern thickness/.store in=\mcThickness, 
pattern thickness = 0.3pt,
pattern radius/.store in=\mcRadius, 
pattern radius = 1pt}
\makeatletter
\pgfutil@ifundefined{pgf@pattern@name@_zbciimwbj}{
\pgfdeclarepatternformonly[\mcThickness,\mcSize]{_zbciimwbj}
{\pgfqpoint{0pt}{0pt}}
{\pgfpoint{\mcSize+\mcThickness}{\mcSize+\mcThickness}}
{\pgfpoint{\mcSize}{\mcSize}}
{
\pgfsetcolor{\tikz@pattern@color}
\pgfsetlinewidth{\mcThickness}
\pgfpathmoveto{\pgfqpoint{0pt}{0pt}}
\pgfpathlineto{\pgfpoint{\mcSize+\mcThickness}{\mcSize+\mcThickness}}
\pgfusepath{stroke}
}}
\makeatother

 
\tikzset{
pattern size/.store in=\mcSize, 
pattern size = 5pt,
pattern thickness/.store in=\mcThickness, 
pattern thickness = 0.3pt,
pattern radius/.store in=\mcRadius, 
pattern radius = 1pt}
\makeatletter
\pgfutil@ifundefined{pgf@pattern@name@_1mz81b1yy}{
\pgfdeclarepatternformonly[\mcThickness,\mcSize]{_1mz81b1yy}
{\pgfqpoint{0pt}{0pt}}
{\pgfpoint{\mcSize+\mcThickness}{\mcSize+\mcThickness}}
{\pgfpoint{\mcSize}{\mcSize}}
{
\pgfsetcolor{\tikz@pattern@color}
\pgfsetlinewidth{\mcThickness}
\pgfpathmoveto{\pgfqpoint{0pt}{0pt}}
\pgfpathlineto{\pgfpoint{\mcSize+\mcThickness}{\mcSize+\mcThickness}}
\pgfusepath{stroke}
}}
\makeatother

 
\tikzset{
pattern size/.store in=\mcSize, 
pattern size = 5pt,
pattern thickness/.store in=\mcThickness, 
pattern thickness = 0.3pt,
pattern radius/.store in=\mcRadius, 
pattern radius = 1pt}
\makeatletter
\pgfutil@ifundefined{pgf@pattern@name@_9pdwee2r5}{
\pgfdeclarepatternformonly[\mcThickness,\mcSize]{_9pdwee2r5}
{\pgfqpoint{0pt}{0pt}}
{\pgfpoint{\mcSize+\mcThickness}{\mcSize+\mcThickness}}
{\pgfpoint{\mcSize}{\mcSize}}
{
\pgfsetcolor{\tikz@pattern@color}
\pgfsetlinewidth{\mcThickness}
\pgfpathmoveto{\pgfqpoint{0pt}{0pt}}
\pgfpathlineto{\pgfpoint{\mcSize+\mcThickness}{\mcSize+\mcThickness}}
\pgfusepath{stroke}
}}
\makeatother
\tikzset{every picture/.style={line width=0.75pt}} 


\draw    (263.5,12693.27) .. controls (263.05,12695.58) and (261.67,12696.52) .. (259.36,12696.07) .. controls (257.05,12695.62) and (255.67,12696.56) .. (255.22,12698.87) .. controls (254.77,12701.18) and (253.38,12702.12) .. (251.07,12701.67) .. controls (248.76,12701.22) and (247.38,12702.16) .. (246.93,12704.47) .. controls (246.48,12706.78) and (245.1,12707.72) .. (242.79,12707.27) .. controls (240.48,12706.82) and (239.1,12707.76) .. (238.65,12710.07) .. controls (238.2,12712.38) and (236.82,12713.32) .. (234.51,12712.87) .. controls (232.2,12712.42) and (230.82,12713.36) .. (230.37,12715.67) .. controls (229.92,12717.99) and (228.54,12718.93) .. (226.22,12718.48) -- (225.83,12718.74) -- (225.83,12718.74) ;
\draw [color={rgb, 255:red, 0; green, 135; blue, 189 }  ,draw opacity=1 ][line width=1.5]    (84.5,12727.52) -- (121.33,12727.74) ;
\draw [color={rgb, 255:red, 208; green, 2; blue, 27 }  ,draw opacity=1 ][line width=1.5]    (84.47,12649.25) -- (262.06,12649.25) ;
\draw    (135.33,12649.47) .. controls (137,12651.13) and (137.01,12652.8) .. (135.35,12654.47) .. controls (133.7,12656.14) and (133.71,12657.81) .. (135.38,12659.47) .. controls (137.05,12661.13) and (137.06,12662.8) .. (135.4,12664.47) .. controls (133.74,12666.14) and (133.75,12667.81) .. (135.42,12669.47) .. controls (137.09,12671.13) and (137.1,12672.8) .. (135.44,12674.47) .. controls (133.78,12676.14) and (133.79,12677.81) .. (135.46,12679.47) .. controls (137.13,12681.13) and (137.14,12682.8) .. (135.48,12684.47) .. controls (133.82,12686.14) and (133.83,12687.81) .. (135.5,12689.47) .. controls (137.17,12691.13) and (137.18,12692.8) .. (135.52,12694.47) .. controls (133.87,12696.14) and (133.88,12697.81) .. (135.55,12699.47) .. controls (137.22,12701.13) and (137.23,12702.8) .. (135.57,12704.47) .. controls (133.91,12706.14) and (133.92,12707.81) .. (135.59,12709.47) .. controls (137.26,12711.13) and (137.27,12712.8) .. (135.61,12714.47) .. controls (133.95,12716.14) and (133.96,12717.81) .. (135.63,12719.47) .. controls (137.3,12721.13) and (137.31,12722.8) .. (135.65,12724.47) -- (135.67,12727.66) -- (135.67,12727.66) ;
\draw  [fill={rgb, 255:red, 255; green, 255; blue, 255 }  ,fill opacity=1 ] (149.67,12649.74) .. controls (149.67,12657.32) and (143.25,12663.47) .. (135.33,12663.47) .. controls (127.42,12663.47) and (121,12657.32) .. (121,12649.74) .. controls (121,12642.16) and (127.42,12636.01) .. (135.33,12636.01) .. controls (143.25,12636.01) and (149.67,12642.16) .. (149.67,12649.74) -- cycle ;
\draw  [pattern=_n879p6llu,pattern size=3.2249999999999996pt,pattern thickness=0.75pt,pattern radius=0pt, pattern color={rgb, 255:red, 0; green, 0; blue, 0}] (149.67,12649.59) .. controls (149.62,12642) and (143.17,12635.89) .. (135.26,12635.93) .. controls (127.34,12635.98) and (120.96,12642.16) .. (121,12649.74) .. controls (121.04,12657.32) and (127.49,12663.44) .. (135.41,12663.39) .. controls (143.33,12663.35) and (149.71,12657.17) .. (149.67,12649.59) -- cycle ;
\draw  [pattern=_izjrheo3s,pattern size=3.2249999999999996pt,pattern thickness=0.75pt,pattern radius=0pt, pattern color={rgb, 255:red, 0; green, 0; blue, 0}] (149.67,12649.39) .. controls (149.62,12641.81) and (143.17,12635.7) .. (135.26,12635.74) .. controls (127.34,12635.78) and (120.96,12641.96) .. (121,12649.55) .. controls (121.04,12657.13) and (127.49,12663.24) .. (135.41,12663.2) .. controls (143.33,12663.16) and (149.71,12656.97) .. (149.67,12649.39) -- cycle ;
\draw  [fill={rgb, 255:red, 255; green, 255; blue, 255 }  ,fill opacity=1 ] (150,12727.74) .. controls (150,12735.32) and (143.58,12741.47) .. (135.67,12741.47) .. controls (127.75,12741.47) and (121.33,12735.32) .. (121.33,12727.74) .. controls (121.33,12720.16) and (127.75,12714.01) .. (135.67,12714.01) .. controls (143.58,12714.01) and (150,12720.16) .. (150,12727.74) -- cycle ;
\draw  [pattern=_vkibh8r0o,pattern size=3.2249999999999996pt,pattern thickness=0.75pt,pattern radius=0pt, pattern color={rgb, 255:red, 0; green, 0; blue, 0}] (150,12727.59) .. controls (149.96,12720) and (143.51,12713.89) .. (135.59,12713.93) .. controls (127.67,12713.98) and (121.29,12720.16) .. (121.33,12727.74) .. controls (121.37,12735.32) and (127.83,12741.44) .. (135.74,12741.39) .. controls (143.66,12741.35) and (150.04,12735.17) .. (150,12727.59) -- cycle ;
\draw  [pattern=_zbciimwbj,pattern size=3.2249999999999996pt,pattern thickness=0.75pt,pattern radius=0pt, pattern color={rgb, 255:red, 0; green, 0; blue, 0}] (150,12727.59) .. controls (149.96,12720) and (143.51,12713.89) .. (135.59,12713.93) .. controls (127.67,12713.98) and (121.29,12720.16) .. (121.33,12727.74) .. controls (121.38,12735.32) and (127.83,12741.44) .. (135.74,12741.39) .. controls (143.66,12741.35) and (150.04,12735.17) .. (150,12727.59) -- cycle ;
\draw [color={rgb, 255:red, 0; green, 135; blue, 189 }  ,draw opacity=1 ][line width=1.5]    (231.33,12727.59) -- (270.5,12727.52) ;
\draw  [fill={rgb, 255:red, 255; green, 255; blue, 255 }  ,fill opacity=1 ] (230.83,12728.74) .. controls (230.83,12736.32) and (224.42,12742.47) .. (216.5,12742.47) .. controls (208.58,12742.47) and (202.17,12736.32) .. (202.17,12728.74) .. controls (202.17,12721.16) and (208.58,12715.01) .. (216.5,12715.01) .. controls (224.42,12715.01) and (230.83,12721.16) .. (230.83,12728.74) -- cycle ;
\draw  [pattern=_1mz81b1yy,pattern size=3.2249999999999996pt,pattern thickness=0.75pt,pattern radius=0pt, pattern color={rgb, 255:red, 0; green, 0; blue, 0}] (230.83,12728.59) .. controls (230.79,12721) and (224.34,12714.89) .. (216.42,12714.93) .. controls (208.51,12714.98) and (202.12,12721.16) .. (202.17,12728.74) .. controls (202.21,12736.32) and (208.66,12742.44) .. (216.58,12742.39) .. controls (224.49,12742.35) and (230.88,12736.17) .. (230.83,12728.59) -- cycle ;
\draw  [pattern=_9pdwee2r5,pattern size=3.2249999999999996pt,pattern thickness=0.75pt,pattern radius=0pt, pattern color={rgb, 255:red, 0; green, 0; blue, 0}] (230.83,12728.59) .. controls (230.79,12721) and (224.34,12714.89) .. (216.42,12714.93) .. controls (208.51,12714.98) and (202.12,12721.16) .. (202.17,12728.74) .. controls (202.21,12736.32) and (208.66,12742.44) .. (216.58,12742.39) .. controls (224.49,12742.35) and (230.88,12736.17) .. (230.83,12728.59) -- cycle ;
\draw [color={rgb, 255:red, 0; green, 135; blue, 189 }  ,draw opacity=1 ][line width=1.5]    (142.5,12715.52) .. controls (145.62,12705.38) and (160.59,12701.85) .. (169.7,12702.48) ;
\draw    (141.5,12741.02) .. controls (143.98,12740.75) and (145.39,12741.78) .. (145.74,12744.11) .. controls (146.01,12746.42) and (147.28,12747.39) .. (149.53,12747.03) .. controls (151.9,12746.78) and (153.21,12747.81) .. (153.45,12750.12) .. controls (153.9,12752.5) and (155.3,12753.4) .. (157.65,12752.81) .. controls (159.7,12751.79) and (161.32,12752.26) .. (162.49,12754.23) .. controls (163.94,12756.15) and (165.58,12756.38) .. (167.42,12754.93) .. controls (169.31,12753.46) and (170.98,12753.63) .. (172.42,12755.44) -- (173,12755.5) ;
\draw    (166.02,12774.82) -- (184,12775) ;
\draw [shift={(163.02,12774.79)}, rotate = 0.58] [fill={rgb, 255:red, 0; green, 0; blue, 0 }  ][line width=0.08]  [draw opacity=0] (5.36,-2.57) -- (0,0) -- (5.36,2.57) -- cycle    ;
\draw    (186.75,12691.5) -- (167.98,12691.31) ;
\draw [shift={(189.75,12691.53)}, rotate = 180.58] [fill={rgb, 255:red, 0; green, 0; blue, 0 }  ][line width=0.08]  [draw opacity=0] (5.36,-2.57) -- (0,0) -- (5.36,2.57) -- cycle    ;
\draw [color={rgb, 255:red, 208; green, 2; blue, 27 }  ,draw opacity=1 ] [dash pattern={on 4.5pt off 4.5pt}]  (176.5,12696.28) -- (175.85,12768.97) ;
\draw [color={rgb, 255:red, 0; green, 135; blue, 189 }  ,draw opacity=1 ][line width=1.5]    (209.67,12716.02) .. controls (206.55,12705.88) and (191.58,12702.35) .. (182.47,12702.98) ;
\draw    (210.67,12741.52) .. controls (210.15,12743.94) and (208.74,12744.97) .. (206.43,12744.6) .. controls (204.12,12744.25) and (202.85,12745.21) .. (202.62,12747.48) .. controls (202.29,12749.84) and (200.94,12750.87) .. (198.57,12750.57) .. controls (196.33,12750.06) and (194.91,12750.89) .. (194.32,12753.06) .. controls (193.17,12755.18) and (191.58,12755.6) .. (189.54,12754.31) .. controls (187.65,12752.9) and (185.92,12753.12) .. (184.37,12754.97) .. controls (182.95,12756.79) and (181.35,12756.95) .. (179.58,12755.44) -- (179,12755.5) ;

\draw (269.56,12639.91) node [anchor=north west][inner sep=0.75pt]    {$p_{1} '$};
\draw (50,12724.93) node [anchor=north west][inner sep=0.75pt]    {$p_{2} ,\sa_2$};
\draw (274.27,12719.54) node [anchor=north west][inner sep=0.75pt]    {$p_{2} ',\sa'_2$};
\draw (66.56,12643.91) node [anchor=north west][inner sep=0.75pt]    {$p_{1}$};
\draw (167.69,12777.8) node [anchor=north west][inner sep=0.75pt]  [font=\normalsize,rotate=-0.76,xslant=-0.06]  {$\ell $};
\draw (144.33,12668.27) node [anchor=north west][inner sep=0.75pt]    {$p_{2} -q_{1} +\ell $};
\draw (267.5,12680.02) node [anchor=north west][inner sep=0.75pt]  [font=\normalsize,rotate=-0.92,xslant=-0.06]  {$k,\eta$};
\draw (-54,12675.5) node [anchor=north west][inner sep=0.75pt]    {$2\text{CCut}_{2,\sa_2}\mathcal{A}_{5,1}^{\sa_2,Q_1 Q_2^4}=$};

\end{tikzpicture}
\]
With this in mind the cut is given  explicitly by 
\[\label{earlycutspin}
\CCut_{2,\sa_2} &\mathcal{A}_{5,1}^{\sa_2,Q_1Q_2^4}=-\frac{4Q_1Q^4_2}{q_1^2} \sum_{\eta,\eta''} p_1\cdot \varepsilon_{\eta}(q_1)\varepsilon^{*\mu}_\eta (q_1) \varepsilon^{*\sigma}_{\eta'}(k)  \int \dd \Phi(-\ell)\, \hat{\delta}(2p_2\cdot(\ell-q_1)) 
\\& \!\!\!\!\times
\varepsilon^{\nu}_{\eta''}(\ell)
\varepsilon^{*\rho}_{\eta''} (\ell)
[\mathcal{J}^{\sa_2}_{\mu\nu}(p_2, -q_1\to p_2-q_1+\ell, -\ell)
\mathcal{J}^{\sa_2}_{\rho\sigma}(p_2 -q_1+\ell, -\ell\to p_2+q_2, k)] \big\rvert_{\sa^2_2},
\]
where the integrand $ \sim[  \mathcal{J}^{\sa_2}_{\mu\nu} \mathcal{J}^{\sa_2}_{\rho\sigma}]\rvert_{\sa^2_2}$ is consistently truncated through spin squared order. One important technical point to note is that what we have done in \eqref{earlycutspin} is to effectively take the classical limit of spinning amplitudes directly on the cut. This is essentially due to the simplicity of the analytic structure of $\mathcal{A}^{\sa_2}_{4,0}$. There, no contributions of the type $\hbar/\hbar$ coming from the interference of ``superclassical$\times$quantum" terms arises, and we can effectively import the classical  spinning variables directly from the product of the two tree-level Compton subamplitudes \eqref{compspin}, where the spin is already classicalised.  This was also noticed and employed in \cite{Akpinar:2025byi}.

The next step is to evaluate the loop integrals. Here, it is enough to observe that these are always of the form 
\[\label{inttt}
I^{\mu_{1}...\mu_{n}}\equiv\int \dd\Phi(-\ell) \hat{\delta}(u_2\cdot (\ell-q_1))\ell^{\mu_1}\cdots\ell^{\mu_n},
\]
which are (tensor) functions of the outgoing photon energy $u_2\cdot k$, as measured from the frame of particle 2.
We find that up to order $\mathcal{O}(\sa^2_2)$ the rank-4 integral  $I^{\mu\nu\rho\sigma}$ is needed at most.\footnote{In the scalar case the integrals    are also of the same type as \eqref{inttt}, but go up to rank-2 at most \cite{Elkhidir:2023dco}.} We further observe that these tensors are indeed very similar to the ones encountered in the earlier section, in the purely classical analysis. The expressions \eqref{inttt} can be solved straightforwardly by reducing onto a Lorentz-invariant basis  and by fixing their coefficients, see for instance \cite{Kosower:2018adc, Elkhidir:2023dco}. We find
\[
&I^0=\frac{u_2\cdot k}{2\pi}\Theta(u_2\cdot k), \,\,\,\,\,I^\mu=\frac{(u_2\cdot k)^2}{2\pi} u_2^\mu\Theta(u_2\cdot k),\,\,\,\,\ I^{\mu\nu}= -\frac{(u_2\cdot k)^3}{6\pi} (\eta^{\mu\nu}-4 \,u_2^\mu u_2^\nu)\Theta(u_2\cdot k),
\\&
\hspace{3cm}I^{\mu\nu\rho}=-\frac{(u_2\cdot k)^4}{6\pi}\left(\frac{1}{2}\eta^{(\mu\nu} u_2^{\rho)}-6 \,u_2^\mu u_2^\nu u_2^\rho\right)  \Theta(u_2\cdot k),\,\,\,\,\ \\& 
\hspace{2cm}I^{\mu\nu\rho\sigma}=\frac{(u_2\cdot k)^5}{30\pi}\left(\frac{1}{8}\eta^{(\mu\nu}\eta^{\rho\sigma)}-\frac{3}{2}\, \eta^{(\mu\nu}u_2^{\rho}u_2^{\sigma)} +48\, u_2^{\mu}u_2^{\nu}u_2^{\rho}u_2^{\sigma} \right) \Theta(u_2\cdot k),
\]
where we have smoothly\footnote{We find no $\epsilon/\epsilon$ terms.} taken now $D=4$.
Note that since  the waveform expression \eqref{eq:far-field-from-waveshape} is on shell and $\omega=u_2\cdot k$ then  $\Theta(u_2\cdot k)=1$.

Putting everything together in the end we find the scalar, spin-orbit and spin squared sectors. Respectively we obtain  
\[\label{cc1}
\CCut_{2,\sa_2} &\mathcal{A}_{5,1}^{\sa_2,Q_1Q_2^4}\bigg\rvert_{\mathcal{O}(\sa^0_2)}=
\frac{2Q_1 Q_2^4m_1 }{3\pi m_2 q_1^2} 
\left(
u_2\cdot k\,u_1\cdot\varepsilon^*(k)
+\gamma\,q_1\cdot\varepsilon^*(k)
\right),
\]

\[\label{cc2}
\CCut_{2,\sa_2} \mathcal{A}_{5,1}^{\sa_2,Q_1Q_2^4}\bigg\rvert_{\mathcal{O}(\sa^1_2)}\!\!\!\!&=
-\frac{2iQ_1 Q_2^4m_1 }{3\pi m_2 q_1^2} 
\Bigg[
(u_2\cdot k)^2\,
\epsilon(\mathsf{a}_2,u_1,\varepsilon^*(k),u_2)
+\gamma\,u_2\cdot k\,
\epsilon(\mathsf{a}_2,q_1,\varepsilon^*(k),u_2)
\\
&\hspace{1.3cm}+\varepsilon^*(k)\cdot q_1\,
\epsilon(\mathsf{a}_2,u_1,q_1,u_2)
-\left(
u_1\cdot k
+\frac{\gamma\,q_1\cdot k}{u_2\cdot k}
\right)
\epsilon(\mathsf{a}_2,\varepsilon^*(k),k,u_2)
\Bigg],
\]

\[\label{cc3}
\\
\CCut_{2,\sa_2} \mathcal{A}_{5,1}^{\sa_2,Q_1Q_2^4}&\bigg\rvert_{\mathcal{O}(\sa^2)}\!\!\!\!\!\!=
\frac{Q_1 Q_2^4m_1}{6\pi m_2q_1^2}\!\Bigg[\!\!
\left(\!
u_1\cdot\varepsilon^*(k)+
\frac{\gamma\,q_1\cdot\varepsilon^*(k)}{u_2\cdot k}
\right)\!\!
\Bigg(\!2\,u_2\cdot k
\left(
(\mathsf{a}_2\cdot q_1)^2\!+(\mathsf{a}_2\cdot k)^2
\right)\!
\\
&\!\!\!\!\!\!+\frac{4\,q_1\cdot k}{u_2\cdot k}\,
\mathsf{a}_2\cdot q_1\,\mathsf{a}_2\cdot k
-\frac{4\mathsf{a}^2_2}{5\,u_2\cdot k}
\left(
5(q_1\cdot k)^2
+10(u_2\cdot k)^2\,q_1\cdot k
+9(u_2\cdot k)^4
\right)
\Bigg)
\\[6pt]
&\!\!\!\!\!\!+
\left(\frac{\gamma\,q_1\cdot\mathsf{a}_2}{u_2\cdot k}
+u_1\cdot\mathsf{a}_2
\right)\!\!
\Bigg(\!\!
-\frac{4\,q_1\cdot k}{u_2\cdot k}\,
\mathsf{a}_2\cdot k\,\varepsilon^*(k)\cdot q_1
\\
&\!\!\!\!\!\!\hspace{2,7cm}
+\frac{4}{5\,u_2\cdot k}
\left(
5(q_1\cdot k)^2
+5(u_2\cdot k)^2\,q_1\cdot k
+12(u_2\cdot k)^4
\right)
\mathsf{a}_2\cdot\varepsilon^*(k)
\Bigg)
\\[6pt]
&\!\!\!\!\!\!+
\left(\frac{\gamma\,q_1\cdot k}{u_2\cdot k}
+u_1\cdot k
\right)\!\!
\Bigg(\!\!
-\frac{4\,q_1\cdot k}{u_2\cdot k}\,
\mathsf{a}_2\cdot\varepsilon^*(k)\,\mathsf{a}_2\cdot q_1
+\frac{4\mathsf{a}^2_2}{u_2\cdot k}\!
\left(
q_1\cdot k+(u_2\cdot k)^2
\right)\!
\varepsilon^*(k)\cdot q_1\!
\Bigg)\!
\Bigg].
\]
Above $\gamma=u_1\cdot u_2$ is  the Lorentz gamma factor, and we choose the gauge $\varepsilon^*\cdot u_2=0$ to simplify the formulae a little \cite{Elkhidir:2023dco,Scheopner:2023rzp}. Note as well that we have evaluated the helicity sums $\sum_{\eta,\eta''}\varepsilon_\eta \varepsilon^*_\eta \varepsilon_{\eta''} \varepsilon^*_{\eta''}$   by inserting the projectors over physical states.  To retrieve the  generic-gauge expressions one  has to simply do the shift
\[
\varepsilon^*_\mu(k) \to \varepsilon^*_\mu (k) -\frac{\varepsilon^*(k)\cdot u_2}{k\cdot u_2} k_\mu,
\]
in each polarization vector inside the CCut expressions. We observe that the scalar part above \eqref{cc1} is the one which was already computed in \cite{Elkhidir:2023dco}. 

Before moving onto the next part, let us briefly comment on the inclusion of spin for particle 1.  If one desires so, we claim that the following replacement is sufficient
\[\label{mmc}
u_1^\mu \to {\exp(i\sa_1*q_1)^\mu}_\nu u_{1}^\nu,
\]
in equations \eqref{eq:CRRW} just as well as in \eqref{cc1}--\eqref{cc3}. The spin operator in \eqref{mmc} is nothing but the matrix exponential introduced  in \eqref{mtrxexponential}, and this substitution is also discussed in section \ref{sub:swaves}. Technically, this is understood from the structure of the calculation. Physically, it is explained by overall mass scaling of the dissipative waveshape, which is  $\mathcal{O}(m_2^{-2})$ (to see this, rescale $p_i^\mu=m_i u_i^\mu$ in eq \eqref{eq:genericWaveshapeIm}). It is indeed known that in the heavy-mass limit the spin of  particle 1 (the heavier one) exponentiates exactly, since the spurious-pole ambiguities drop out \cite{Siemonsen:2019dsu, Menezes:2022tcs,Bianchi:2023lrg}. Having understood this, we continue to adopt $\sa^\mu_1\equiv0$.

\paragraph{Comparison to the results of section \ref{sec: crr}}
    
We end this section by commenting on the equivalence between the radiation-reaction waveshapes, first obtained using classical and then with quantum field theory methods. Comparing the two expressions for the waveshape, \eqref{eq:genericWaveshapeIm} and \eqref{classwshap}, we observe that their equality requires that
\[
\;\mathrm{CCut}_{2,\mathsf a_2}\mathcal A^{\mathsf a_2,Q_1Q_2^4}_{5,1}
=-\frac{4i\,m_1m_2}{q_1^2}\;\mathcal W_{\rm RR}.
\]
Taking this overall normalisation into account and comparing directly the spin sectors in  \eqref{eq:CRRW} to the ones in \eqref{cc1}--\eqref{cc3}, written in terms of scalar products of four-vectors one can observe a direct match term by term. To see this one can substitute the explicit form  of $n^\mu=u_1^\mu+\gamma\, q_1^\mu/u_2\cdot k$   inside the classically-derived waveshape expressions \eqref{eq:CRRW}. Equivalently, one can project both expressions onto a  particular frame and establish their agreement. A convenient choice is
\cite{Akpinar:2024meg}
\[
&k^\mu = \omega_1 \hat k^\mu = \omega_1 (1, \sin \theta \cos \phi ,\sin \theta \sin \phi, \cos \theta), \\
&u_1^\mu = (1,0,0,0),\,\,\,u_2^\mu = (\gamma,0,0,\gamma\beta), \\
&\varepsilon^{\mu}_\pm (k) = \frac{1}{\sqrt{2}}(\partial_\theta  \hat k^\mu \pm \frac{i}{\sin \theta} \partial_\phi \hat k^\mu ), \\
& q_1^\mu = (0, q_1^x,q_1^y , \frac{\gamma \omega_1 - \omega_1 \gamma \beta \cos \theta}{\gamma \beta}),
\\&
\sa^\mu_2=(\beta \sa^z_2 ,\sa^x_2,\sa^y_2, \sa^z_2),
\]
where $\omega_1=u_1\cdot k$ is the photon energy in the rest frame of particle 1, so that $\omega=u_2\cdot k=\gamma\omega_1(1-\beta\cos\theta)$, and the kinematic vectors obey their on-shell identities and the spin was constrained by $\sa_2\cdot u_2=0$. Comparing onto a common frame also has the virtue of trivializing any possible difference between  two results that can be traced back to  Bianchi identities $\partial^{[\rho}F^{\mu\nu]}=0\approx\partial^{[\rho}{}^*F^{\mu\nu]}$ or Schoutening of the type $v_i^\mu v_j^\nu v_k^\rho v_r^\sigma \epsilon_{[\mu\nu\rho\sigma}v_{s\lambda]} v^{\lambda}_p=0$, where $v^\mu_i=\{\varepsilon^\mu_\eta, \,\,q^\mu_1, \,\,k^\mu, \,\, u_{1,2}^\mu,\sa^\mu_2\}$.

In the next section we will conclude our paper with a discussion of radiation reaction for spinning bodies in general relativity and future research directions.

\section{Discussion and outlook}\label{sec:concle}
Let us begin by  commenting  on the equivalence we have established in this work. First, inspired by an old article of Sidney Coleman \cite{RM-2820-PR} we derived a new version of the ALD radiation reaction equation that is able to encompass dissipation of a spinning body up to order $\mathcal{O}(\sa^2)$. This is a classical derivation that makes extensive use of dimensional regularisation to renormalise short-scale divergences of the radiative self-fields. Tensor reduction and integration are immediate and lead to the equations of motion \eqref{eq:CRR}. We solve these perturbatively, starting from the straight line approximation \cite{Elkhidir:2023dco} all the way to the  $\sim\mathcal{O}(\sa_2^2 Q_1 Q_2^4)$  dissipative EM spinning waveshape. This is given by  the expressions \eqref{classwshap}  and \eqref{eq:CRRW}. Both the equations in \eqref{eq:CRR} and the dissipative one-loop EM waveform, computed here through spin-squared order, represent new results.

Then, we moved onto an on-shell derivation of the same results, which allowed us to confirm and test our framework. Building upon the results of  \cite{Elkhidir:2023dco}, where radiation  reaction was rephrased in terms of the imaginary part of the one-loop Compton amplitude, we derive the same dissipative contributions using products of tree-level Compton amplitudes. Radiation reaction is here determined by a product of two tree-level Compton amplitudes where one internal messenger goes on-shell. As the two calculations are  independent, this provides an important consistency check on our method. 

Additionally, another non-trivial confirmation of the methods of section \ref{sec: crr} is detailed  in appendix \ref{appendix subl}, where we show that non-Schott terms can also be derived. These match the previous results of \cite{Jakobsen:2023tvm} and extend them to spin-squared order.

Let us now comment on possible future avenues. 
One  research direction lies in using the dissipative equations to try to constrain classical Compton amplitudes. In fact, we stress that the formalism provided here can in principle be pushed to arbitrarily high-spin multipoles. The only roadblock is again the knowledge of the tree-level classical Compton amplitude, which in the spinning case presents spurious poles for $\mathcal{O}(\sa^{n>2})$. These classical poles also affect the dynamical mass function $\mathcal{M}$ of \cite{Scheopner:2023rzp}. Then, our methods can help test future Compton-amplitude proposals via the computation of dissipative observables in a new way with \eqref{eq:CRR}. In fact,  this type of analysis has mainly been done via conservative observables so far. There, one would simply parametrise additional spinning operators \cite{deFirmian:2026mln, Siemonsen:2019dsu},  compute observables and compare \cite{Bautista:2024agp}. We also underline the fact that our off-shell equations  are in principle valid for arbitrary trajectories, so  in this sense having \eqref{eq:CRR} does not necessarily restrict us to scattering. Additionally, one can push the calculations of dissipative observables to higher orders. This could involve, for instance,  a computation of the two-loop EM waveform, spinning energy losses \cite{Herrmann:2021tct} or higher loops for the deflection observables  \cite{Menezes:2022tcs} and spin kicks \cite{Luna:2023uwd, Maybee:2019jus,Haiashi:2026eyn}. Another option is instead to study spinning radiation reaction in higher dimensions, building upon the recent results of \cite{Campanella:2026wqt,Cangemi:2026cjy,Gambino:2024uge}.  Furthermore, it would also be interesting to add $O(F^2)$ tidal operators in our analysis, and to see whether the $\sqrt{\text{Kerr}}$/Kerr
worldsheet structure first proposed in \cite{Guevara:2020xjx} (See also \cite{Kim:2024grz})  is broken or preserved by the structure of the dissipative equations \eqref{eq:CRR}.

A final, and perhaps the most important, extension is represented by the obvious inclusion of gravity to our analysis. This is also motivated by the recent impressive progress in the computation of post-Minkowskian  (PM) observables with QFT and amplitude methods, see \cite{Bjerrum-Bohr:2022blt,Buonanno:2022pgc} for a review on the topic as well as recent state of the art results for the PM momentum deflection \cite{Klemm:2024wtd,Driesse:2024xad,Driesse:2024feo,Driesse:2026qiz,Bern:2025wyd, Bern:2026oqp} and waveforms \cite{DeAngelis:2023lvf, Herderschee:2023fxh,Georgoudis:2023lgf,Georgoudis:2025vkk,Brandhuber:2023hhy,Brunello:2024ibk,Brunello:2025eso,Bohnenblust:2023qmy,Bohnenblust:2025gir,DeAngelis:2026kew}.  In general relativity, one way to understand the motion of bodies and their radiation reaction is through the celebrated ``MiSaTaQuWa" equation \cite{Mino:1996nk, Quinn:1996am}, describing the linear gravitational self-force. Extensions with spin of this equation are not known to the best of our knowledge. The MiSaTaQuWa equation can be understood as the gravity analogue of the ALD equation, except it is more complicated. This is due to nonlocal gravitational tails: the gravitational Green's function  has support both \emph{on} and \emph{inside} the lightcone and, to determine the trajectory and momentum of a particle, one needs to integrate this equation over the whole history of the particle \cite{Pound:2021qin, Poisson:2011nh,Havas:1957qve,Havas:1962zz}. Understanding the spinning extension of the MiSaTaQuWa equation, with cuts and dimreg, is the topic of future work \cite{future}. Inspired by the analysis carried out in this paper, we will establish a connection between the spin methods of sections \ref{sec: crr} and  \ref{sec:ccut} towards a new formulation of the spinning self-force equations. While supplying the gravitational CCut operator with  spin is fairly straightforward  -- as above, one can simply insert tree-level spinning amplitudes in the cut -- we can already anticipate where the classical analysis will lead. Naturally, in gravity the object that is to be regularised is the Christoffel symbol (and its derivatives). This will play a similar role to Maxwell's tensor in \eqref{rrf}. In momentum space this can be expressed as 
\begin{equation}
\partial_{\alpha_1}\cdots\partial_{\alpha_n}\Gamma^\mu_{\alpha\beta}(z(0))=\frac{m\,\kappa^2(-i)^{n+1}}{2}\!\!\!\int\!\hd^Dk \!\int_{-\infty}^0\!\!\mathrm{d}\tau\, \frac{e^{ik\cdot z(\tau)}}{k^2+i\epsilon k^0}k_{\alpha_1}\cdots k_{\alpha_n}\!\left(k_{(\alpha}{H_{\beta)}}^\mu(k)-k^\mu H_{\alpha\beta}(k)\right),
\end{equation}
where $H^{\mu\nu}(k)$ is the   Fourier transform of the metric, defined by
\begin{equation}\label{eq:hmunuSolution}
    h^{\mu\nu}(z(0))=m \,\kappa\int\! \hd^Dk\int_{-\infty}^0\!\mathrm{d}\tau\, \frac{e^{ik\cdot z(\tau)}}{k^2+i\epsilon k^0}H^{\mu\nu}(k)\ ,
\end{equation}
where we are gathering details about spin into  the explicit definition of $H^{\mu\nu}(k)$. The explicit expression of the metric field is easily found following for instance \cite{Vines:2017hyw, Arkani-Hamed:2019ymq, Menezes:2022tcs,Gambino:2024uge,Damgaard:2022jem,Damgaard:2026ocb,Campanella:2026wqt,Bianchi:2023lrg}. Note also that in gravity nonlinearities arise as the metric above $H_{\mu\nu}$ is sourced both by masses and by the gravitational field itself \cite{Scheopner:2023rzp}. For instance, we anticipate here that the ``QED"-like analysis of \ref{sec: crr} leads to a new spinning Schott term, coming from a regularised Christoffel symbol of the form 
\begin{align}\label{gamma}
    \Gamma^{\mu\, (1)}_{\alpha\beta} &= \frac{1}{6}Jm\kappa^2 \left(\dddot{u}_{(\alpha}{\epsilon_{\beta)}}^{\mu}(\sa,u)+u_{(\alpha}{\epsilon_{\beta)}}^{\mu}(\sa,\dddot{u})+u_{(\alpha}{\epsilon_{\beta)}}^{\mu}(\dddot{\sa},u)-3 u_\alpha u_\beta \epsilon^\mu(\sa,u,\dddot{u})\right) + \mathcal O(\kappa^3)\ ,
\end{align}
where we omitted nonlinear terms coming from graviton bulk interactions.  Interestingly, we report that \eqref{gamma} leads to a gravitational Schott term that is not a pure-gauge contribution for the dissipative waveform as obtained from the regularised MPD equations \cite{Mathisson:1937zz,Papapetrou:1951pa,Dixon:1970zza}. This already shows both the importance and applicability of our analysis. 
We then leave these  exciting research directions to future work.

\section*{Acknowledgments}
We thank Dogan Akpinar, Zvi Bern, Chris de Firmian, Donal O'Connell, Anton Ilderton, Michael Saavedra, Trevor Scheopner, Maarten van de Meent  and Fei Teng  for discussions and useful comments. We are especially grateful to Dogan Akpinar for helpful conversations and collaboration during the initial stages of this work. M.S. has been
supported in part by the European Research Council under Advanced Investigator grant ERC–AdG–885414, by the U.S. Department of Energy (DOE) under award number DE-SC0009937, and by the European Research Council (ERC) Horizon Synergy Grant “Making Sense of the Unexpected in the Gravitational Wave Sky” grant
agreement no. GWSky–101167314. M.S. also acknowledges support from the Mani L. Bhaumik Institute for Theoretical Physics. 
The work of P.V.M. is supported by a Science and Technology Facilities Council (STFC) quota studentship.
No new data were generated or analysed
during this study.
Views and opinions expressed are however those of the authors only and do not necessarily reflect those of the European Union or European Research Council Executive Agency.
Neither the European Union nor the granting authority can be held responsible for them.

\appendix
 
\section{Subleading dissipative fields}\label{appendix subl}
The leading order radiation reaction equations take the general form of sums of terms of the form $Q_2^2 \,X^{(n)}\cdot Y f(u,\sa)$, where $X^{(n)}$ denotes either $u_2$ or $\sa_2$ with any number of time derivatives. This means that they provide $\mathcal O(Q_1Q_2^3)$ corrections to the motion, owing to the fact that both $u_2^{(n)},\sa_2^{(n)}\sim \mathcal O(Q_1Q_2)$ at leading order. However, the CRR method can be done keeping higher order terms of the form $Q_2^2\, X^{(n)}\cdot Y^{(m)}f(u_2,\sa_2)$ or $Q_2^2\, (X^{(n)}\cdot Y)(Z^{(m)}\cdot W)f(u_2,\sa_2)$, scaling as $\mathcal O(Q_1^2Q_2^4)$ at leading order. We thus report here the non-Schott results 
\begin{align} 
    \dot{p}^\mu_2\big|_{{\text{RR}},\text{NLO}}&= \frac{Q_2^2}{6\pi}\dot{u}_2^2u_2^\mu+\frac{Q_2^2}{6\pi}\left(\epsilon^\mu(u_2,\dot{\sa}_2,\ddot{u}_2)-\epsilon^\mu(\sa_2,\dot{u}_2,\ddot{u}_2)-2u_2^\mu\epsilon(\sa_2,u_2,\dot{u}_2,\ddot{u}_2)\right)\nn\\
    &\quad - \frac{Q_2^2}{120\pi}\Bigl( 5\bigl(11\dot\sa_2\cdot \dddot u_2 + 16\ddot\sa_2\cdot\ddot u_2+10\dddot\sa_2\cdot\dot u_2\bigr)\sa_2^\mu +15\sa_2\cdot\dddot u_2\dot\sa_2^\mu + 20\sa_2\cdot \ddot u_2\ddot\sa_2^\mu\nn\\
    &\qquad +10\sa_2\cdot\dot u_2\dddot\sa_2^\mu - \bigl(10(\sa_2\cdot\ddot u_2)^2+\sa_2^2(49\dot u_2\cdot \dddot u_2+33\dot u_2^2)\bigr)u_2^\mu - 60\sa_2\cdot\dddot\sa_2 \dot u_2^\mu\nn\\
    &\qquad -60\sa_2\cdot\ddot\sa_2 \ddot u_2^\mu - 30\sa_2\cdot\dot\sa_2 \dddot u_2^\mu\Bigr)\ ,
    \label{eq:NLOEOMp}\\
    \dot\sa_2^\mu\big|_{{\text{RR}},\text{NLO}}&= -\frac{Q_2^2}{12\pi m_2}\Bigl(6\bigl(\epsilon^\mu(\sa_2,\dot\sa_2,\ddot u_2)+\epsilon^\mu(\sa_2,\ddot\sa_2,\dot u_2)+u_2^\mu(\epsilon(\sa_2,u_2,\dot\sa_2,\ddot u_2)+\epsilon(\sa_2,u_2,\ddot\sa_2,\dot u_2))\bigr)
    \nn\\
    &\qquad-\sa_2\cdot\ddot u_2\epsilon^\mu(\sa_2,u_2,\dot u_2)+\sa_2\cdot\dot u_2\epsilon^\mu(\sa_2,u_2,\ddot u_2)\Bigr).
\end{align}
These contributions supplement \eqref{eq:CRR}. They are not total-derivative terms and thus contribute to the impulse at NNLO (two loops). The linear in spin contributions to the radiation reaction equation of motion for $u$ were computed in \cite{Jakobsen:2023tvm}, and we find agreement with their results. There the author found, specialising to the  Landé  factor $g_L=2$, with no external fields, and translated to our conventions:
\begin{equation}\label{eq:GusRNLO}
    m\dot{u}^{\mu\, (1)}_{2}\big|_{{\text{RR}},\text{NLO}} = -\frac{Q_2^2}{6\pi}\epsilon^{\mu}(\ddot{u}_2,\dot{\sa}_2,u_2)\ .
\end{equation}
We cannot compare directly to \eqref{eq:NLOEOMp}, since $p^\mu=\mathcal{M}\ph^\mu$; the $\mathcal O(\sa)$ difference between $\mathcal{M}$ and $m$ is important. Instead, it can be shown that
\begin{equation}
    \frac{\dd \ph^\mu}{\dd\tau} = \frac{1}{\mathcal{M}}P^{\mu\nu}\frac{\dd p_\nu}{\dd\tau}\ ,
\end{equation}
and at linear order in spin $\ph^\mu$ and $u^\mu$ coincide (their difference \eqref{eq:hidden} is $\mathcal O(\sa^2)$). Applying this to \eqref{eq:NLOEOMp}, and combining with a suitable Schouten identity, shows agreement with \eqref{eq:GusRNLO}. In particular, the relevant Schouten identity is
\begin{align}
    0=u_2^{[\mu}\epsilon^{\nu\rho\sigma\tau]}\sa_{2\nu}\dot{u}_{2\rho}\ddot{u}_{2\sigma} u_{2\tau} &= u_2^\mu \epsilon(\sa_2,u_2,\dot{u}_2,\ddot{u}_2)+\dot{u}_2^2\epsilon^\mu(\sa_2,\dot{u}_2,u_2)+\epsilon^\mu(\sa_2,\dot{u}_2,\ddot{u}_2)\nn\\
    &= u_2^\mu \epsilon(\sa_2,u_2,\dot{u}_2,\ddot{u}_2)+\epsilon^\mu(\sa_2,\dot{u}_2,\ddot{u}_2) + \mathcal O(Q^6)
\end{align}
using the fact that $u_2\cdot \sa_2=0$, $u_2\cdot\dot{u}_2=0$ and $u_2\cdot \ddot{u}_2=-\dot{u}_2^2\sim \mathcal O(Q^4)$.
 
\bibliographystyle{JHEP}
\bibliography{ALDandMiSaTaQuwa} 

\end{document}